%% file: main.tex
\documentclass[11pt,a4paper]{article}
\pdfoutput=1
\usepackage[utf8]{inputenc}
\usepackage[T1]{fontenc}
\usepackage{jheppub}

\usepackage{bbold}
\usepackage[extdef]{delimset}
\usepackage{mathtools}
\usepackage{physics}

\usepackage{glossaries}
\glsdisablehyper
\usepackage{enumitem}
\setlist{itemsep=1pt,topsep=1pt,parsep=1pt,partopsep=1pt}
\usepackage{tikz-feynman}
\usetikzlibrary{arrows}
\usetikzlibrary{cd}
\usepackage{subcaption}

\usepackage{cleveref}
\usepackage[numbers,sort&compress]{natbib}

\input{commands}

\input{glossaries}

\title{Momentum mappings for subtractions at higher orders in QCD}
\author[a,1]{Vittorio Del Duca%
\note{On leave from INFN LNF, Italy.},}
\author[a]{Nicolas Deutschmann,}
\author[a,b,2]{Simone Lionetti%
\note{On leave from IPPP Durham, United Kingdom.}}
\affiliation[a]{Institute for Theoretical Physics, ETH Zurich\\
Wolfgang-Pauli-Str.\ 27, 8093 Z\"urich, Switzerland}
\affiliation[b]{Department of Information Technologies, Hochschule Luzern\\
Suurstoffi 1, 6343 Rotkreuz, Switzerland}
\emailAdd{delducav@itp.phys.ethz.ch, ndeutschmann@itp.phys.ethz.ch} 
\emailAdd{simone.lionetti@hslu.ch}
\arxivnumber{}

\abstract{
Subtraction schemes provide a systematic way to compute fully-differential cross sections beyond the leading order in the strong coupling constant.
These methods make singular real-emission corrections integrable in phase space by the addition of suitable counterterms.
Such counterterms may be defined using momentum mappings, which are parametrisations of the phase space that factorise the variables that describe the particles becoming unresolved in some infrared or collinear limit from the variables that describe an on-shell phase space for the resolved particles.
In this work, we review existing momentum mappings in a unified framework and introduce new ones for final-collinear and soft counterterms.
The new mappings work in the presence of massive particles and with an arbitrary number of soft particles or of clusters of collinear particles,
making them fit for subtraction methods at any order in perturbation theory.
The new mapping for final-collinear counterterms is also used to elucidate relations among existing final-collinear mappings.
}

\begin{document}

	\maketitle
	
	\input{introduction}
	\input{general}

	\input{finalcoll}
	\input{soft}
	\input{independence}
	\input{conclusion}
	\section*{Acknowledgements}
	We would like to thank Valentin Hirschi for many useful discussions as well as for his collaboration on related projects.
	This  project  has  received  funding  from  the  European  Research  Council~(ERC)  under  grant  agreement  No  694712  (PertQCD).
	The work of SL was supported by ETH Z\"urich and by an STSM Grant from the COST Action CA16201 PARTICLEFACE.
	\appendix
	\input{lorentz_transformations}
	\input{massive_lorentz}

	\bibliographystyle{JHEP}
	\bibliography{biblio}

\end{document}

%% file: commands.tex
\let\abs\physabs
\let\comm\physcomm
\DeclareMathDelimiterSet{\abs}[1]{\selectdelim[l]|#1\selectdelim[r]|}
\DeclareMathDelimiterSet{\comm}[2]{%
\selectdelim[l][#1\selectdelim[c],#2\selectdelim[r]]}
\DeclareMathDelimiterSet{\intd}[1]{%
\selectdelim[l]\langle#1\selectdelim[r]\rangle}

\let\Tr\physTr
\NewDocumentCommand{\Tr}{so}{%
\mathrm{Tr}\IfValueT{#2}{\IfBooleanTF#1{\brk[s]*{#2}}{\brk[s]{#2}}}}
\let\Re\physRe
\NewDocumentCommand{\Re}{so}{%
\mathrm{Re}\IfValueT{#2}{\IfBooleanTF#1{\brk[s]*{#2}}{\brk[s]{#2}}}}
\let\Im\physIm
\NewDocumentCommand{\Im}{so}{%
\mathrm{Im}\IfValueT{#2}{\IfBooleanTF#1{\brk[s]*{#2}}{\brk[s]{#2}}}}

\NewDocumentCommand{\lcpar   }{m o m}{#1_{\IfValueT{#2}{#2}\parallel #3}}
\NewDocumentCommand{\lcperp  }{m o}{#1_{\IfValueT{#2}{#2}\perp}}
\NewDocumentCommand{\lctrans }{m o}{#1_{\IfValueT{#2}{#2}\mathrm{T}}}
\NewDocumentCommand{\lccplus }{m o}{#1{}_{\IfValueT{#2}{#2}}^+{}}
\NewDocumentCommand{\lccminus}{m o}{#1{}_{\IfValueT{#2}{#2}}^-{}}
\NewDocumentCommand{\lccperp }{m o}{#1{}_{\IfValueT{#2}{#2}}^\perp{}}

\NewDocumentCommand{\jetF}{m o}{J^{(#1)}\IfValueT{#2}{_{#2}}}
\NewDocumentCommand{\pdf}{m o}{f_{#1\IfValueT{#2}{/#2}}}

\NewDocumentCommand{\lm}{m o}{\mathbf{#1}\IfValueT{#2}{_{#2}}}
\NewDocumentCommand{\ct}{m o}{\mathrm{#1}\IfValueT{#2}{\brk{#2}}}
\NewDocumentCommand{\cur}{m o}{\mathcal{#1}\IfValueT{#2}{_{#2}}}

\newcommand{\resc}{\kappa}

\newcommand{\madgraph}{\texttt{MG5\_aMC}}
\newcommand{\madnklo}{\texttt{MadNkLO}}

\newcommand{\colorful}{\textsc{CoLoRFul}}

\def\beq{\begin{equation}}
\def\eeq{\end{equation}}
\def\beeq{\begin{eqnarray}}
\def\eeeq{\end{eqnarray}}

\newcommand\nn         {\nonumber}

\newcommand{\ti}[1]    {\tilde{#1}}

\NewDocumentCommand{\PS}{o}{\dd{\Phi\IfValueT{#1}{_{#1}}}}

\makeatletter
\newcommand{\vect}[1]{\vec{#1} \@ifnextchar{^}{\hspace{0.1em}}{}}
\makeatother

%% file: glossaries.tex
\newacronym{RHS}{RHS}{right-hand side}
\newacronym{LHS}{LHS}{left-hand side}

\newacronym{QFT}{QFT}{Quantum Field Theory}
\newacronym{QCD}{QCD}{Quantum Chromodynamics}
\newacronym{SM}{SM}{Standard Model}
\newacronym{EW}{EW}{electroweak}
\newacronym{SCET}{SCET}{soft-collinear effective theory}

\newacronym{LHC}{LHC}{Large Hadron Collider}
\newacronym{LEP}{LEP}{Large Electron-Positron collider}

\newacronym{DIS}{DIS}{deep-inelastic scattering}

\newacronym{LO}{LO}{leading order}
\newacronym{NLO}{NLO}{next-to-leading order}
\newacronym{NNLO}{NNLO}{next-to-next-to-leading order}
\newacronym{N3LO}{N\ensuremath{{}^3}LO}{next-to-next-to-next-to-leading order}
\newacronym{LL}{LL}{leading logarithmic}
\newacronym{NLL}{NLL}{next-to-leading logarithmic}
\newacronym{NNLL}{NNLL}{next-to-next-to-leading logarithmic}
\newacronym{N3LL}{N\ensuremath{{}^3}LL}
	{next-to-next-to-next-to-leading logarithmic}

\newacronym{R}{R}{real}
\newacronym{V}{V}{virtual}
\newacronym{RR}{RR}{double-real}
\newacronym{RV}{RV}{real-virtual}
\newacronym{VV}{VV}{double-virtual}

\newacronym{KLN}{KLN}{Kinoshita--Lee--Nauenberg}
\newacronym{DGLAP}{DGLAP}{Dokshitzer--Gribov--Lipatov--Altarelli--Parisi}
\newacronym{CS}{CS}{Catani--Seymour}
\newacronym{FKS}{FKS}{Frixione--Kunszt--Signer}
\newacronym{BRST}{BRST}{Becchi--Rouet--Stora--Tyutin}
\newacronym{BPHZ}{BPHZ}{Bogoliubov--Parasiuk--Hepp--Zimmermann}

\newacronym{IRC}{IRC}{infrared and collinear}
\newacronym{UV}{UV}{ultraviolet}

\newacronym{XS}{XS}{cross section}
\newacronym{PDF}{PDF}{parton distribution function}
\newacronym{BR}{BR}{branching ratio}

\newacronym{OPE}{OPE}{operator product expansion}
\newacronym{IBP}{IBP}{integration-by-parts}
\newacronym{MI}{MI}{master integral}
\newacronym{CDR}{CDR}{conventional dimensional regularisation}

%% file: introduction.tex
\section{Introduction}
\label{sec:intro}

The discovery of the Higgs boson~\cite{Aad:2012tfa,Chatrchyan:2012xdj} at the CERN \gls{LHC} and the subsequent study of its properties has been the crowning achievement of the \gls{SM}.
These results fixed the value of the last unknown parameter of the \gls{SM} and confirmed many predictions of the properties of the Higgs boson, such as the value of its coupling strengths to the \gls{SM} fermions.
While the \gls{LHC}, as a hadron collider, was designed as a discovery machine, it is now also our best tool to improve the precision of our knowledge of the \acrlong{SM}, both in the Higgs sector, where we have much to learn, and in the other aspects of the theory.
As statistics are steadily increasing and experimental techniques improving, the experimental uncertainties are progressively shrinking and projections for the end of the \gls{LHC} programme indicate that it will provide precision results in the Higgs sector as well as on the strong and electroweak interactions.
In many cases, in particular for Higgs physics, it is expected that the leading source of uncertainty on \gls{SM} observables will be due to theoretical predictions.
As a result, it is crucial that we improve the precision of theoretical calculations, which can often be achieved by going to higher orders in perturbation theory.

The \gls{NLO} both in the \gls{QCD} and in the \gls{EW} couplings is now routinely achievable for most processes thanks to the development of highly efficient automated tools.
Since in many cases of interest \gls{NLO} predictions are not sufficient to match the experimental precision, the next target are \gls{NNLO} calculations in the \gls{QCD} coupling for which no fully automated method exists yet.

A major issue in performing higher-order perturbative calculations in gauge theories such as \gls{QCD} is the treatment of \gls{IRC} divergences.
It is known that these singularities must vanish from properly defined infrared-safe observables by virtue of \gls{KLN} theorem~\cite{Kinoshita:1962ur,Lee:1964is} and that they manifest themselves through the universal factorisation of amplitudes,
which is known fully to the second order in the perturbative expansion in \gls{QCD},
for the double-virtual~\cite{Catani:1998bh,Sterman:2002qn,Aybat:2006wq,Aybat:2006mz},
real-virtual~\cite{Bern:1994zx,Bern:1998sc,Kosower:1999rx,Bern:1999ry,Catani:2000pi}
and double-real corrections~\cite{Campbell:1997hg,Catani:1999ss,DelDuca:1999iql,Kosower:2002su},
with several contributions being known also to the third order~\cite{Almelid:2015jia,Bern:2004cz,Badger:2004uk,Duhr:2013msa,Li:2013lsa,Duhr:2014nda,Catani:2003vu,Badger:2015cxa,DelDuca:1999iql,Birthwright:2005ak,Duhr:2006,Catani:2019nqv}. 
However, translating this understanding into systematic ways to perform calculations is a daunting task.

At the \gls{NLO}, exploration of this \gls{IRC} behavior of amplitudes has led to the establishment of subtraction methods~\cite{Frixione:1995ms,Catani:1996vz} as the standard approach, in which real-emission corrections, which exhibit \gls{IRC} divergences when integrated over phase space, are made integrable by the addition of suitable counterterms.
These counterterms exhibit the same divergent behaviour as real-emission matrix elements, but are typically much simpler functions of the kinematics.
This permits to integrate them in dimensional regularisation, expose their singularities as poles in the regulator and cancel them against the \gls{IRC} poles of the virtual loop corrections - as predicted by the \gls{KLN} theorem.

At \gls{NNLO}, a number of approaches for handling real \gls{IRC} singularities have been successfully used to perform calculations.
Fully-differential results have been obtained in processes that involve up to three particles in the final state for colour-neutral initial states and up to two particles in the final state for hadron collisions. 
These calculations are based on either slicing schemes,
like $q_T$~\cite{Catani:2007vq}
or $N$-jettiness~\cite{Boughezal:2015dva,Gaunt:2015pea},
or subtraction schemes, like for instance antenna~\cite{
GehrmannDeRidder:2005cm,Daleo:2006xa,GehrmannDeRidder:2005aw,
GehrmannDeRidder:2005hi,Daleo:2009yj,Gehrmann:2011wi,Boughezal:2010mc,
GehrmannDeRidder:2012ja,Currie:2013vh},
\colorful~\cite{Somogyi:2005xz,Somogyi:2006da,
Somogyi:2006db,Somogyi:2008fc,Aglietti:2008fe,Somogyi:2009ri,
Bolzoni:2009ye,Bolzoni:2010bt,DelDuca:2013kw,Somogyi:2013yk}, 
residue-improved~\cite{Czakon:2010td,Czakon:2011ve,Czakon:2014oma,Czakon:2019tmo}, 
nested soft-collinear~\cite{Caola:2017dug,Caola:2018pxp,Delto:2019asp,Caola:2019nzf,Caola:2019pfz}, 
projection-to-Born~\cite{Cacciari:2015jma} subtraction.

All the subtraction schemes mentioned above are based on devising a set of counterterms which approximate the matrix element in the limits where they become singular, such that the difference be computable in four dimensions and the counterterms can be integrated in $d$-dimensions as a Laurent series in the regulator.
Defining a scheme amounts to specifying how these counterterms are obtained for a generic process, or at least a generic class of processes.
There is of course a lot of freedom in the selection of counterterms, as is exhibited by the number of well-established schemes and by the continued appearance of new approaches~\cite{Magnea:2018hab,Magnea:2018ebr,Herzog:2018ily}.

All the existing schemes exploit the factorisation properties of amplitudes in the singular limits, which are the points in phase space where some massless partons are collinear to one another or soft.
The leading behaviour of the amplitude close to the singular surfaces in phase space is known, and in order to achieve a working subtraction it must be the case that the sum of all counterterms features the same leading behaviour.
Counterterms, however, are functions of phase space as a whole (possibly equal to zero in parts of it in the case of sector-based approaches) and one must define them away from the singular limits.
This is where schemes differ from one another, since the definition of their singular limits is far from enough to make the counterterms unique.

One aspect of the freedom of choice in the definition of counterterms for subtraction schemes is momentum mappings.
Momentum mappings are parametrisations of the phase space where the variables that describe the particles becoming unresolved in some infrared or collinear limit are factorised from the variables that describe an on-shell phase space for the resolved particles.
This factorisation property is key to make a subtraction method general using the following procedure:
\begin{itemize}
	\item for a given limit, we choose a mapping that factorises phase space into a lower-multiplicity phase space and ``unresolved variables'';
	\item we write our counterterm as a universal singular factor (e.g.\ an Altarelli--Parisi kernel) multiplied by a squared amplitude with the appropriate reduced multiplicity;
	\item we choose the momenta of the lower-multiplicity phase space as the arguments of the lower-multiplicity squared amplitude and we express the universal singular factor in terms of the ``unresolved variables'' and possibly of the lower-multiplicity phase-space momenta.
\end{itemize}

The factorisation of both the integrand and the parametrisation allows then the $d$-dimensional integration of the singular factor over the ``unresolved variables'' for fixed lower-multiplicity kinematics.
Thanks to this procedure, the integral of the real-emission contribution yields a Laurent series in the dimensional regulator, whose poles cancel the poles of the corresponding virtual squared amplitudes \emph{locally} in the lower-multiplicity phase space.

Momentum mappings were introduced at \gls{NLO} where the original dipole subtraction~\cite{Catani:1996vz} uses a factorising parametrisation for massless partons, of which that of the \colorful\ subtraction~\cite{Somogyi:2006cz} is a variation.
A variant of the mapping of Ref.~\cite{Catani:1996vz} was introduced to handle massive particles~\cite{Catani:2002hc} and another solution was proposed subsequently by Nagy and Soper~\cite{Nagy:2007ty}.
As we shall see in \cref{sec:review}, different schemes use different mappings at \gls{NNLO}.

Because momentum mappings are always a means to an end, their properties and defining features have been little studied so far.
However, the current activity in setting up subtraction schemes that are truly general calls for a transversal study of momentum mappings.
The aim of the present paper is to make a step in this direction.
More precisely, we will review a number of existing momentum mappings in a unified framework, introduce new ones and present important observations about their application.

The paper is organised as follows.
\Cref{sec:review} provides a more detailed discussion of what a momentum mapping is by setting up explicitly their definition at \gls{NLO} for collinear and soft configurations and showing how they are realised in different existing schemes both at \gls{NLO} and \gls{NNLO}.
\Cref{sec:fcoll} introduces a new momentum mapping for final-collinear counterterms which is shown to be a generalisation of existing mappings.
The new momentum mapping works in the presence of massive particles and with an arbitrary number of clusters of collinear particles.
It can be used to show the equivalence of the mappings of Refs.~\cite{Catani:2002hc} and \cite{Nagy:2007ty} in the case of massive particles
with a single recoiler.
\Cref{sec:softsect} introduces a new momentum mapping for soft counterterms which works in the presence of massive particles, as well as
in specific kinematic configurations where existing soft mappings fail.
\Cref{sec:independence} presents ideas to allow subtraction schemes to be setup in a way where integrated counterterms can be computed once and for all independently of the choice of mappings, which we realise explicitly on the specific case of final state \gls{NLO} collinear counterterms.
\Cref{sec:conclusion} presents our conclusions.

%% file: general.tex
\section{Review of momentum mappings}
\label{sec:review}

This section is dedicated to introducing momentum mappings with explicit examples, setting up definitions and providing a resource where the different approaches used in the literature and their properties are described.
We will first review the example of \colorful\ subtraction for final-state \gls{NLO} singularities, which is built on both a soft and a collinear mapping, allowing us to illustrate both important aspects and possible issues relating to mappings.
We then move on to discuss the three main different types of mappings and their realisations in different subtraction schemes.
Finally we discuss how these elementary mappings can be combined to handle counterterms where disjoint sets of particles become unresolved.

\subsection{An example at NLO: \colorful\ subtraction}
\label{sec:colorful_sub}

As a first example, let us describe how \colorful\ subtraction handles the regularisation of the real-emission corrections to the process $e^+\,e^-\to q\,\bar{q}$ where $q$ is a massless quark, i.e.\ $e^+\;e^-\to q\;\bar{q}\;g$.
The matrix element for this process diverges in three limits: when the gluon momentum becomes zero ($p_g\to 0$), or when it becomes collinear to the momentum of either the quark ($p_g\parallel p_q$) or the anti-quark ($p_g\parallel p_{\bar{q}}$).
The well-known collinear and soft factorisation formulae for the squared amplitude multiplied by an \gls{IRC}-safe observable $\mathcal{O}$ read as follows,
\begin{align}
\label{eq:qqgsoftlimit}
&\lim_{p_g\to 0}\left|{\cal M}_{q\bar{q}g}(p_q, p_{\bar{q}},p_g) \right|^2{\cal O}\left(p_q, p_{\bar{q}},p_g\right)
\nonumber\\&\hspace{4em}=
-8\pi\alpha_s \mu^{2\epsilon} C_F \frac{p_q \cdot p_{\bar{q}} }{p_g\cdot p_q\;p_g\cdot p_{\bar{q}}} \left|{\cal M}_{q\bar{q}}(p_q, p_{\bar{q}}) \right|^2 {\cal O}\left(p_q, p_{\bar{q}}\right)\\
\label{eq:qqgcolllimit}
&\lim_{p_g \parallel p_q}\left|{\cal M}_{q\bar{q}g}(p_q, p_{\bar{q}},p_g) \right|^2{\cal O}\left(p_q, p_{\bar{q}},p_g\right)
\nonumber\\&\hspace{4em}=
8\pi\alpha_s \mu^{2\epsilon}  \frac{C_F}{2\,p_g\cdot p_q} \frac{1+(1-z)^2}{z} \left|{\cal M}_{q\bar{q}}\left(p_q+p_g, p_{\bar{q}}\right) \right|^2{\cal O}\left(p_q+p_g, p_{\bar{q}}\right)\\
&\lim_{p_g \parallel p_{\bar{q}}}\left|{\cal M}_{q\bar{q}g}(p_q, p_{\bar{q}},p_g) \right|^2{\cal O}\left(p_q, p_{\bar{q}},p_g\right)
\nonumber\\&\hspace{4em}=
8\pi\alpha_s \mu^{2\epsilon}  \frac{C_F}{2\,p_g\cdot p_{\bar{q}}} \frac{1+(1-z')^2}{z'} \left|{\cal M}_{q\bar{q}}\left(p_q, p_{\bar{q}}+p_g\right) \right|^2{\cal O}\left(p_q, p_{\bar{q}}+p_g\right),
\end{align}
where $z p_q = (1-z) p_g$ and $z' p_{\bar{q}} = (1-z') p_g$ in the respective collinear limits, $\mu$ is the regularisation scale, $\epsilon=(d-4)/2$ is the dimensional regulator, $\alpha_s$ is the strong coupling and $C_F$ is the fundamental Casimir of the strong interaction gauge group.
The amplitudes ${\cal M}_{q\bar{q}}$ and  ${\cal M}_{q\bar{q}g}$ are respectively the amplitudes for the Born ($e^+\;e^-\to q\;\bar{q}$) and real-emission ($e^+\;e^-\to q\;\bar{q}\;g$) processes.
Note that amplitudes are unambiguous only when their arguments are momentum conserving and on-shell, meaning that both $\left|{\cal M}_{q\bar{q}}(p_q, p_{\bar{q}}) \right|^2$ and $\left|{\cal M}_{q\bar{q}}(p_q+p_g, p_{\bar{q}}) \right|^2$ are well-defined only in the appropriate limits, as momentum conservation between the initial and final state would otherwise not be respected in the former and the on-shell condition for the quark would be violated in the latter.
As we discussed in the introduction, a momentum mapping resolves this ambiguity by defining \emph{mapped momenta} $\tilde p_q$ and $\tilde{p}_{\bar{q}}$ as functions of $p_q$, $p_{\bar{q}}$ and $p_g$ such that:
\begin{enumerate}[label=($\roman*$)]
\item \label{enum:mapprop:1} they are on-shell and momentum conserving outside of the strict limit;
\item \label{enum:mapprop:2} the integration measure is factorised.
\end{enumerate}
Next, we illustrate how this works.

\subsubsection{Soft mapping}
\label{sec:softexample}

The issue with using the right-hand side of \cref{eq:qqgsoftlimit} to define a soft counterterm is that the momenta $p_q$ and $p_{\bar{q}}$ do not add up to the initial state total momentum $Q = p_{e^+}+p_{e^-}$, such that the matrix element is not well-defined outside of the strict limit.
The solution proposed in the \colorful\ scheme is to use instead the momenta,
\begin{align}
\tilde{p}_q^\mu = \Lambda^\mu{}_\nu[\lambda Q, Q-p_g] \frac{p_q^\nu}{\lambda},
\qquad
\tilde{p}_{\bar{q}}^\mu = \Lambda^\mu{}_\nu[\lambda Q, Q-p_g] \frac{p_{\bar{q}}^\nu}{\lambda},
\end{align}
where
\begin{equation}
\label{eq:softexample:lambdadef}
\lambda = \sqrt{\frac{(Q-p_g)^2}{Q^2}},
\end{equation}
is the ratio of centre-of-mass energy of the $q\bar{q}$ system to the one of the $q\bar{q}g$ system, and $\Lambda[\lambda Q, Q-p_g]$ is a Lorentz transformation that maps $Q-p_g$ to $\lambda Q$, given in \cref{eq:lorentzlambda}.
The role of $\Lambda$ is easy to understand in the rest frame of $Q$, where it ensures that $\tilde p_q+\tilde p_{\bar{q}}$ has a null total 3-momentum.
Replacing $ p_{q/\bar{q}}$ with $\tilde p_{q/\bar{q}}$ does not affect the leading behaviour of the right-hand side of \cref{eq:qqgsoftlimit}, since in the soft-gluon limit $p_g\to 0$ we have $\lambda\to 1$ and therefore $\tilde p_{q/\bar{q}}\to p_{q/\bar{q}}$.
As a result, a valid soft counterterm is
\begin{align}
{\cal M}\!S_g^{q\bar{q}}(p_q,p_{\bar{q}},p_g) &= -8\pi\alpha_s \mu^{2\epsilon} C_F \frac{\tilde p_q \cdot \tilde p_{\bar{q}} }{p_g\cdot \tilde p_q\;p_g\cdot \tilde p_{\bar{q}}} \left|{\cal M}_{q\bar{q}}(\tilde p_q, \tilde p_{\bar{q}}) \right|^2 {\cal O}\brk*{\tilde{p}_q, \tilde{p}_{\bar{q}}}.
\end{align}
Note that the momenta in the eikonal factor are taken to be the mapped momenta.
This is an arbitrary choice, since choosing to keep the original quark momenta in the eikonal factor would again yield the same leading behaviour in the soft limit.
This counterterm provides an appropriate regulation for the soft divergence of the matrix element, as it is a well defined function of the same variables as the matrix element and reproduces its behaviour in the soft limit.
In the computation of the local counterterm required for integrating the real-emission matrix element over the 4-dimensional phase space, property \ref{enum:mapprop:1} is exploited: the arguments of the reduced matrix element $\left|{\cal M}_{q\bar{q}}(\tilde p_q, \tilde p_{\bar{q}}) \right|^2$ are on-shell momenta that verify momentum conservation, which allows it to be unequivocally defined and efficiently derived.

Property \ref{enum:mapprop:2} relates to the $d$-dimensional integration of the counterterm, which is performed to expose the poles in the analytic regulator at the local level in the Born phase space.
In the phase-space integral that defines the cross section, the \colorful\ soft mapping provides a change of variables that allows us to factorise the integration measure as follows,
\begin{align}
\int \dd{\Phi_{q\bar{q}g}} = \int & \frac{\dd[d]{p_q}}{(2\pi)^{d-1}}\delta_+\brk*{p_q^2}\frac{\dd[d]{p_{\bar{q}}}}{(2\pi)^{d-1}}\delta_+\brk*{p_{\bar{q}}^2}\nonumber\\
&\times\frac{\dd[d]{p_g}}{(2\pi)^{d-1}}\delta_+\brk*{p_g^2}(2\pi)^d \delta^d(Q-p_q-p_{\bar{q}}-p_g) \nonumber\\
= \int & \frac{\dd[d]{\tilde p_q}}{(2\pi)^{d-1}}\delta_+\brk*{\lambda^2\tilde p_q^2} \frac{\dd[d]{\tilde p_{\bar{q}}}}{(2\pi)^{d-1}}\delta_+\brk*{\lambda^2\tilde p_{\bar{q}}^2}\lambda^{2d} \nonumber\\
&\times \frac{\dd[d]{p_g}}{(2\pi)^{d-1}}\delta_+\brk*{p_g^2}(2\pi)^d\delta^d\left(\lambda \Lambda^{-1}\left(Q - \tilde p_q - \tilde p_{\bar q} \right)\right)\nonumber\\
= \int & \dd{\Phi_{\tilde{q}\tilde{\bar{q}}}}\,
\lambda^{d-4}\frac{\dd[d]{p_g}}{(2\pi)^{d-1}}\delta_+\brk*{p_g^2}\theta\brk*{\lambda} \theta\brk*{1- \lambda}.
\label{eq:softphasespace}
\end{align}
Using this expression, we can now write the $d$-dimensional integral of the counterterm,
\begin{align}
\avg*{{\cal M}\!S_g^{q\bar{q}}}&=-8\pi\alpha_s \mu^{2\epsilon} C_F\,
\int \dd{\Phi_{\tilde{q}\tilde{\bar{q}}}} \left|{\cal M}_{q\bar{q}}(\tilde p_q, \tilde p_{\bar{q}}) \right|^2{\cal O}\brk*{\tilde{p}_q,\tilde{p}_{\bar{q}}}\nonumber\\
&\qquad \times \int \frac{\dd[d]{p_g}}{(2\pi)^{d-1}}\delta_+\brk*{p_g^2}\lambda^{d-4}\theta\brk*{\lambda} \theta\brk*{1- \lambda} \frac{\tilde p_q \cdot \tilde p_{\bar{q}} }{p_g\cdot \tilde p_q\;p_g\cdot \tilde p_{\bar{q}}}.
\end{align}
The universal integrated counterterm is the integral over the unresolved gluon momentum of the eikonal factor,
\begin{align}
\avg*{S_g^{q\bar{q}}}\left(\tilde{p}_q,\tilde{p}_{\bar{q} }, Q\right)
= \int \frac{\dd[d]{p_g}}{(2\pi)^{d-1}}\delta_+\brk*{p_g^2}\lambda^{d-4}\theta\brk*{\lambda} \theta\brk*{1- \lambda} \frac{\tilde p_q \cdot \tilde p_{\bar{q}}}{p_g\cdot \tilde p_q\;p_g\cdot \tilde p_{\bar{q}}}.
\end{align}
Note that the dependence on $Q$ indicated on the left-hand side is not explicit in the formula, but is generated by $\lambda$ defined through \cref{eq:softexample:lambdadef}.
The integral in the last equation can be done analytically once and for all in $d$-dimensions and it exposes the implicit soft singularities of the real-emission matrix element as explicit poles in the dimensional regulator.
This leaves the integral over the Born phase space, which is usually done numerically as the integrand is now integrable in the limit $d\to 4$ and the observable function can be arbitrarily complicated,
\begin{align}
\avg*{{\cal M}\!S_g^{q\bar{q}}}&=-8\pi\alpha_s \mu^{2\epsilon} C_F
\times\int \dd{\Phi_{q\bar{q}}} \left|{\cal M}_{q\bar{q}}(\tilde p_q, \tilde p_{\bar{q}}) \right|^2{\cal O}\brk*{\tilde{p}_q,\tilde{p}_{\bar{q}}}\avg*{S_g^{q\bar{q}}}\brk*{\tilde{p}_q,\tilde{p}_{\bar{q}},Q}.
\label{eq:colorfulsoftintegrated}
\end{align}
Furthermore, the singular eikonal factor for a soft gluon emitted from a quark-antiquark dipole is universal and the phase-space factorisation used above is generalisable to arbitrary final states as we will illustrate below.
As a result, the integrated singular factor $\avg*{S_g^{q\bar{q}}}$ will appear unchanged for arbitrarily complicated processes that feature this type of singularity.

\subsubsection{Collinear mapping}

The discussion of the \colorful\ approach for the subtraction of the \gls{NLO} soft divergences already outlined the main advantages of using momentum mappings for subtraction: they allow for the definition of reduced kinematics which completely separate the variables which control a singular limit of the matrix element from the variables that describe the reduced process, i.e.\ the variables that enter the matrix element and the observable.
This in turns permits the analytic integration of the singular factor independently of the process and the observable.

This is of course also true for collinear singularities and their associated mappings.
Let us consider the limit $g\parallel q$ in the example at hand; the issue is that the momentum of the parent quark $p_q+p_g$ which appears on the right-hand side of \cref{eq:qqgcolllimit} is on-shell only in the exact limit.
The collinear mapping used in \colorful\ defines a valid parent momentum via a ``democratic'' shift proportional to the total momentum $Q$ of the process, and adjusts all other (massless) momenta through rescaling,
\begin{gather}
\tilde p_{qg}^\mu = \frac{1}{1-\alpha}\left(p_q^\mu+p_g^\mu-\alpha Q^\mu\right),\qquad
\tilde{p}_{\bar{q}}^\mu = \frac{1}{1-\alpha}p_{\bar{q}}^\mu,\nonumber\\
\alpha = \frac{1}{2}\left(y_{(gq)Q}- \sqrt{y_{(gq)Q}^2-4y_{gq}}\right),
\label{eq:colorrescal}
\end{gather}
where $y_{ab}=2 p_a \cdot p_b /Q^2$, $y_{aQ}=2 p_a \cdot Q /Q^2$ and $y_{(ab)Q}=2 (p_a+p_b) \cdot Q /Q^2$.
As in the case of the soft mapping, an essential feature of this mapping is that the transformation between $p_{qg},p_{\bar{q}}$ and $\tilde p_{qg},\tilde p_{\bar{q}}$ becomes trivial in the collinear limit.
Indeed in this limit $y_{gq} \to 0$, therefore $\alpha\to 0$ which in turn implies $\tilde{p}_{qg}\to p_{qg}$ and $\tilde{p}_{\bar{q}}\to p_{\bar q}$.
One can then define a local counterterm that is valid over all of phase space and reproduces the collinear limit of the matrix element as
\begin{align}
{\cal M}\!C_{gq}^{q\bar{q}}(p_q,p_{\bar{q}},p_g) &= 8\pi\alpha_s \mu^{2\epsilon} \frac{C_F}{2\,p_g\cdot p_q} \frac{1+(1-z)^2}{z} \left|{\cal M}_{q\bar{q}}\brk*{\tilde p_{qg}, \tilde p_{\bar{q}}} \right|^2{\cal O}\brk*{\tilde p_{qg}, \tilde p_{\bar{q}}}.
\end{align}
The real-emission phase space can be rewritten as a convolution over $s_{qg}$,
\begin{align}
\int \dd{\Phi_{q\bar{q}g}}
&=\int \frac{\dd[d]{p_q}}{(2\pi)^{d-1}}\delta_+\brk*{p_q^2} \frac{\dd[d]{p_{\bar{q}}}}{(2\pi)^{d-1}}\delta_+\brk*{p_{\bar{q}}^2} \frac{\dd[d]{p_g}}{(2\pi)^{d-1}}\delta_+\brk*{p_g^2}
	(2\pi)^d \delta^d(Q-p_q-p_{\bar{q}}-p_g)\nonumber\\
&=\int_0^{Q^2}\!\frac{\dd{s_{qg}}}{2\pi}
	\int \frac{\dd[d]{p_{qg}}}{(2\pi)^{d-1}}\delta_+\brk*{p_{qg}^2-s_{qg}}\frac{\dd[d]{p_{\bar{q}}}}{(2\pi)^{d-1}}\delta_+\brk*{p_{\bar{q}}^2} (2\pi)^d \delta^d(Q-p_{qg}-p_{\bar{q}})\nonumber\\
&\phantom{=\int}\times\int \frac{\dd[d]{p_g}}{(2\pi)^{d-1}}\delta_+\brk*{p_g^2}\frac{\dd[d]{p_q}}{(2\pi)^{d-1}}\delta_+\brk*{p_q^2}(2\pi)^d \delta^d(p_{qg}-p_q-p_g),
\end{align}
and using the mapping one finds
\begin{align}
\int \dd{\Phi_{q\bar{q}g}}
&=\int \frac{\dd[d]{\tilde{p}_{qg}}}{(2\pi)^{d-1}}\delta_+\brk*{\tilde p_{qg}^2} \frac{\dd[d]{\tilde{p}_{\bar{q}}}}{(2\pi)^{d-1}}\delta_+\brk*{\tilde p_{\bar{q}}^2}
	(2\pi)^d \delta^d(Q-\tilde p_{qg}-\tilde p_{\bar{q}})
	\int_0^{Q^2}\!\frac{\dd{s_{qg}}}{2\pi}J\brk*{s_{qg},\Phi_{\widetilde{qg}\tilde{\bar{q}}}}\nonumber\\
&\phantom{=\int}\times \int \frac{\dd[d]{p_g}}{(2\pi)^{d-1}}\delta_+\brk*{p_g^2}\frac{\dd[d]{p_q}}{(2\pi)^{d-1}}\delta_+\brk*{p_q^2}(2\pi)^d \delta^d(p_{qg}-p_q-p_g)\nonumber\\
&=\int \dd{\Phi_{\widetilde{qg}\tilde{\bar{q}}}} \int_0^{Q^2}\!\frac{\dd{s_{qg}}}{2\pi}J\left(s_{qg},\Phi_{\widetilde{qg}\tilde{\bar{q}}}\right)\int \dd{\Phi_{qg}\left(s_{qg},\Phi_{\widetilde{qg}\tilde{\bar{q}}}\right)},
\label{eq:psdecayfact}
\end{align}
where $J\left(s_{qg},\Phi_{\widetilde{qg}\tilde{\bar{q}}}\right)$ is the Jacobian of the change of variable $(p_{qg},p_{\bar{q}})\to(\tilde p_{qg},\tilde p_{\bar{q}})$ at fixed $s_{qg}$.
As in the soft case, we can then separate the integral over the reduced phase-space variables $\Phi_{\widetilde{qg}\tilde{\bar{q}}}$ that enter the matrix element and observable in the counterterm from the integral of the singular factor taken over the decay phase space of the off-shell parent quark with momentum $p_{qg}^2=s_{qg}$ to the quark-gluon final state, $\Phi_{qg}$.
This decay phase space depends on the momentum of the parent particle $p_{qg}$, which is itself a function of $s_{qg}$ and $\Phi_{\widetilde{qg}\tilde{\bar q}}$ through the mapping, which we explicit in the integral measure above.
This phase-space factorisation yields the integrated counterterm,
\begin{align}
\avg*{{\cal M}\!C_{gq}^{q\bar{q}}} &= 8\pi\alpha_s \mu^{2\epsilon}  C_F\int \dd{\Phi_{\widetilde{qg}\tilde{\bar{q}}}} \left|{\cal M}_{q\bar{q}}\brk*{\tilde p_{qg}, \tilde p_{\bar{q}}} \right|^2{\cal O}\brk*{\tilde p_{qg}, \tilde p_{\bar{q}}} \nonumber\\
&\times\int_0^{Q^2}\! \frac{\dd{s_{qg}}}{s_{qg}}\frac{J\left(s_{qg},\Phi_{\widetilde{qg}\tilde{\bar{q}}}\right)}{2\pi}  \int \dd{\Phi_{qg} \frac{1+(1-z)^2}{z}},
\end{align}
which, as in the case of the soft counterterm, features the universal integrated singular factor,
\begin{align}
\avg*{C_{gq}}\brk*{\tilde p_{qg},Q} = \int_0^{Q^2}\!\frac{\dd{s_{qg}}}{s_{qg}}\frac{J\left(s_{qg},\Phi_{\widetilde{qg}\tilde{\bar{q}}}\right)}{2\pi} \int \dd{\Phi_{qg} \frac{1+(1-z)^2}{z}}.
\end{align}
The fact that this integral only depends on $\tilde p_{qg}$ and $Q$ is specific of the mapping adopted for this scheme, and in general the dependence can be over the whole reduced phase space.
This integral can be performed once and for all to expose the related IRC phase-space singularities as poles in the dimensional regulator, leaving the reduced phase-space integral,
\begin{align}
\avg*{{\cal M}\!C_{gq}^{q\bar{q}}} &= 8\pi\alpha_s \mu^{2\epsilon}  C_F\int \dd{\Phi_{\widetilde{qg}\tilde{\bar{q}}}} \left|{\cal M}_{q\bar{q}}\brk*{\tilde p_{qg}, \tilde p_{\bar{q}}} \right|^2{\cal O}\brk*{\tilde p_{qg}, \tilde p_{\bar{q}}}\avg*{C_{gq}}\brk*{\tilde p_{qg},Q},
\end{align}
to be done numerically for arbitrary observables after it has been combined with the virtual contribution.

Now that we have illustrated the role that mappings play in subtractions, let us review the different choices that have been made in the existing methods for \gls{NLO} and \gls{NNLO} subtractions.
We will first focus on mappings used to define counterterms that regulate final-collinear limits, then those associated to initial-collinear limits and finally those used to regulate soft limits.%
\footnote{
Note that in this work we distinguish \emph{final-collinear} configurations, where a cluster of particles in the final state are collinear to each other,
\emph{initial-collinear} singularities, where a set of particles in the final state is collinear to an initial-state parton,
and \emph{soft} singularities, where the momenta of some particles vanish.
This is in contrast to the language of dipole and antenna subtraction, where one considers \emph{final-final}, \emph{initial-final}, and \emph{initial-initial} emissions
according to the pair of legs that radiate extra particles.
}

\subsection{Final-collinear mappings}

The basis of all the final-collinear mappings we present in this section is the well-known factorisation of arbitrary phase spaces into the production of a parent particle and its decay, already used in \cref{eq:psdecayfact}.
Let us consider $n = m+k-1$ final-state particles with momenta $\{p\}_{n} = \{p_1, \ldots, p_{n}\}$.
Their phase space may be factorised as follows,
\begin{align}
\int \dd{\Phi_{n}(\{p\}_{n};Q)}&=\int \brk[s]*{\prod_{i=1}^{n} \frac{\dd[d]{p_i}}{(2\pi)^{d-1}}\delta_+\brk*{p_i^2-m_i^2}} (2\pi)^d \delta^d\brk3{Q-\sum_{i=1}^{n}p_i}\nonumber\\[1em]
&=\int \frac{\dd{s_K}}{2\pi}\int \brk[s]*{\prod_{i=1}^{m-1} \frac{\dd[d]{p_i}}{(2\pi)^{d-1}}\delta_+\brk*{p_i^2-m_i^2}} \frac{\dd[d]{p_K}}{(2\pi)^{d-1}}\delta_+\brk*{p_K^2-s_K}\nonumber\\
&\qquad\times(2\pi)^d \delta^d\brk3{Q-\sum_{i=1}^{m-1} p_i-p_K}\nonumber\\
&\qquad\times \int \brk[s]*{\prod_{r=m}^{n} \frac{\dd[d]{p_r}}{(2\pi)^{d-1}}\delta_+\brk*{p_r^2-m_r^2}} (2\pi)^d \delta^d\brk3{p_K-\sum_{r=m}^{n}p_r}\nonumber\\[1em]
&=\int_{s_\text{min}}^{s_0} \frac{\dd{s_K}}{2\pi}\int \dd{\Phi_{m}(p_1,\dots,p_{m-1},p_K;Q)} \int \dd{\Phi_{k}(p_m,\ldots,p_n;p_K)},
\label{eq:decayfactorisation}
\end{align}
where
\begin{align}
	\sqrt{s_\text{min}}=\sum_{r=m}^n m_r, \qquad \sqrt{s_0} = Q - \sum_{i=1}^{m-1} m_i.
\end{align}

What we call a final-collinear mapping is a change of variables $\{p_1,\dots,p_{m-1},p_K\}$ $\to$ $\{\tilde p\}_m = \{\tilde p_1,\dots,\tilde p_{m-1}, \tilde p_K\}$ such that the $\{\tilde p\}_m$ span the phase space of $m$ particles with total momentum $Q$ and fixed ($s_K$-independent) masses $\set{\tilde{m}_1, \ldots, \tilde{m}_{m-1}, \tilde{m}_K}$.
In practice, in order to regulate the divergences associated to $k$ final state particles becoming collinear, we are interested in keeping $m_i$ unchanged for $i=1,\dots,m-1$ and setting $\tilde m_K=0$ so that $p_K$ describes the momentum of an on-shell massless parton.
The existence of a singularity in the $(m \dots n)$-collinear limit also depends on having $m_{m}=\dots=m_{n}=0$.

In this section we will always consider that all momenta not involved in the collinear splitting are mapped.
It is however of course always possible to leave some momenta unchanged and to affect only a subset through the mapping.
When discussing this possibility, also in the context of initial-collinear and soft mappings, we will refer to the momenta that are affected by the mapping as \emph{recoilers} while we will refer to those that are not changed as \emph{spectators}.

\subsubsection{Dipole mapping}
In the dipole subtraction scheme for \gls{NLO} subtractions~\cite{Catani:1996vz}, the following mapping is used to define an on-shell reduced matrix element for the $(s,t)$-collinear limit: we specify a massless recoiler momentum, which we take here to be $p_r$, and define
\begin{align}
\tilde p_{st}^\mu &= p_{st}^\mu - \frac{s_{st}}{2 p_{st}\cdot p_r} p_r^\mu,&
\tilde p_r^\mu &= \left(1+\frac{s_{st}}{2 p_{st}\cdot p_r}\right)p_r^\mu,&
\tilde p_i^\mu &= p_i^\mu \quad\text{for}\quad i\neq r,s,t,
\end{align}
with $p_{a_1\ldots a_n} = p_{a_1}+\ldots + p_{a_n}$.
It is easy to see that momentum conservation is obtained since we have $p_r+p_s+p_t = \tilde p_r + \tilde p_{st}$.
One can easily generalise this mapping to $k$ unresolved particles $(t_1,\dots, t_k)$\footnote{The mapping with $k=2$ is used in antenna 
subtraction~\cite{Kosower:2002su,Gehrmann-DeRidder:2003pne,GehrmannDeRidder:2005cm}.},
\begin{align}
&\tilde p_{t_1\dots t_k}^\mu = p_{t_1\dots t_k}^\mu - \frac{s_{t_1\dots t_k}}{2 p_{t_1\dots t_k}\cdot p_r} p_r^\mu, \qquad
\tilde p_r^\mu = \left(1+\frac{s_{t_1\dots t_k}}{2 p_{t_1\dots t_k}\cdot p_r}\right)p_r^\mu,\nonumber\\
&\tilde p_i^\mu = p_i^\mu \quad\text{for}\quad i\neq r,t_1,\ldots,t_k.
\end{align}
The phase space for $n = m+k-1$ final-state particles is written explicitly as
\begin{align}
\int \dd{\Phi_{n}(\{p\}_{n};Q)} &= \int \dd{\Phi_m(\{\tilde p\}_m;Q)} \int_0^{\tilde s }\frac{\dd{s_{t_1\dots t_k}}}{2\pi} \left(1-\frac{s_{t_1\dots t_k}}{\tilde s}\right)^{d-3}\int \dd{\Phi_k\left(p_{t_1},\ldots,p_{t_k};p_{t_1\dots t_k}\right)},
\end{align}
where $\tilde s= 2 \tilde p_{t_1\dots t_k} \cdot \tilde p_{r}$.
The term $(1-s_{t_1\dots t_k}/\tilde s)^{d-3}$ is the Jacobian of the change of variable $(p_{r},p_{t_1\dots t_k})\to (\tilde p_{r},\tilde p_{t_1\dots t_k})$.

\subsubsection{Rescaling mapping}
\label{sec:rescalingmapping}
The rescaling mapping, which is used in the \colorful\ scheme, can be seen as a generalisation of the dipole mapping where the pair of collinear momenta $(s,t)$ recoils against all other final-state particles instead of a single recoiler $r$.
In order for the rescaling mapping to be applicable all recoilers need to be massless.
Indeed, each of the resolved momenta is rescaled by an appropriate factor to restore momentum conservation as shown in \cref{eq:colorrescal} and below,
\begin{align}
& \tilde p_{st}^\mu = \frac{1}{1-\alpha_{st}} \brk{p_{st}^\mu - \alpha_{st} Q^\mu},\qquad
  \tilde p_i^\mu = \frac{1}{1-\alpha_{st}} p_i^\mu \quad\text{for}\quad i\neq s,t,\nonumber\\
&\text{where}\quad \alpha_{st}= \frac{1}{2}\left(y_{(st)Q} - \sqrt{y_{(st)Q}^2-4 y_{st}}\right),
\end{align}
It is again easy to generalise this to multiple unresolved momenta $t_1, \dots, t_k$~\cite{Somogyi:2006da} as follows,
\begin{align}
& \tilde p_{t_1\dots t_k}^\mu = \frac{1}{1-\alpha_{t_1\dots t_k}} \brk{p_{t_1\dots t_k}^\mu - \alpha_{t_1\dots t_k} Q^\mu},\qquad
  \tilde p_i^\mu = \frac{1}{1-\alpha_{t_1\dots t_k}}p_i^\mu \quad\text{for}\quad i\neq t_1,\dots,t_k,\nonumber\\
&\text{where}\quad \alpha_{t_1\dots t_k}= \frac{1}{2}\left(y_{(t_1\dots t_k)Q} - \sqrt{y_{(t_1\dots t_k)Q}^2-4 y_{t_1\dots t_k}}\right).
\label{eq:rescalingmapping}
\end{align}
Incidentally, we note that in the centre-of-mass frame $\alpha_{t_1\dots t_k}Q$ is the anti-collinear (minus) component of $p_{t_1\dots t_k}$ in a light-cone parametrisation along the direction defined by $\tilde{p}_{t_1\dots t_k}$ or equivalently by $p_{t_1\dots t_k}$ itself, i.e.
\begin{equation}
	\alpha_{t_1\dots t_k}Q = p_{t_1\dots t_k} \cdot \binom{1}{\vec{p}_{t_1\dots t_k}/\abs{\vec{p}_{t_1\dots t_k}}}.
\end{equation}
The relation in \cref{eq:rescalingmapping} is invertible and allows us to express $\alpha_{t_1\dots t_k}$ in terms of $s_{t_1\dots t_k}$ and $\tilde p_{t_1\dots t_k}$ as
\begin{equation}
	\alpha_{t_1\dots t_k} = \frac{\sqrt{\tilde{y}^2+4 y_{t_1\dots t_k}(1-\tilde{y})}-\tilde{y}}{2 \left(1-\tilde{y}\right)},
\end{equation}
where $\tilde y = 2 \tilde p_{t_1\dots t_k}\cdot Q/Q^2$ and $y_{t_1\dots t_k} = s_{t_1\dots t_k}/Q^2$. We can then write the following phase space factorisation formula for $n = m+k-1$ final-state particles,
\begin{align}
& \int \dd{\Phi_{n}(\{p\}_{n};Q)} \\
&= \int \dd{\Phi_m(\{\tilde p\}_m;Q)} \int_0^{Q^2}\frac{\dd{s_{t_1\dots t_k}}}{2\pi} \frac{\tilde y(1-\alpha_{t_1\dots t_k})^{(m-1)(d-2)-1}}{\tilde y +2 \alpha_{t_1\dots t_k} (1-\tilde y) } \int 
\dd{\Phi_k\left(\left\{p_{t_i}\right\}_k;p_{t_1\dots t_k}\right)}, \nn
\end{align}
where the term $[\tilde{y}(1-\alpha_{t_1\dots t_k})^{(m-1)(d-2)-1}]/[\tilde y +2 \alpha_{t_1\dots t_k} (1-\tilde y)]$ is the Jacobian of the change of variables
$\{p_{1},\dots,p_{m-1},p_{t_1\dots t_k}\}$ $\to$ $\{\tilde p_{1},\dots,\tilde p_{m-1},\tilde p_{t_1\dots t_k}\}$.
In this case, a parametrisation that leads to a simpler expression for the phase space is the one already adopted in the \colorful\ scheme,
\begin{align}
\label{eq:rescalingmappingPS}
& \int \dd{\Phi_{n}(\{p\}_{n};Q)} \\
&= \int \dd{\Phi_m(\{\tilde p\}_m;Q)} \frac{\tilde y Q^2}{2\pi} \int_0^{1} \dd{\alpha_{t_1\dots t_k}} (1-\alpha_{t_1\dots t_k})^{(m-1)(d-2)-1} \int 
\dd{\Phi_k\brk*{\{p_t\}_k;p_{t_1\dots t_k}}}. \nn
\end{align}

\subsubsection{Lorentz mapping}
\label{sec:lorentzmapping}
An issue with the rescaling mapping defined above is that the resolved momenta need to be massless so that the rescaling does not change their mass.
This holds both for the parent of the collinear particles and for the other final-state recoilers.
A quick fix to apply the rescaling mapping in the presence of massive final-state particles would be to recoil only against the massless ones.
Depending on the process this is however not always possible, as in the case of the real-emission contribution $e^+\,e^- \to t\,\bar{t}\,g\,g$ in the limit where the two gluons become collinear.
An option that is applicable in this case, proposed by Nagy and Soper~\cite{Nagy:2007ty}, is to restore momentum conservation in the reduced phase space using a Lorentz transformation.
In the case of two unresolved massless momenta $s$ and $t$ it takes the form,
\begin{equation}
\begin{aligned}
& \tilde p_{st}^\mu = \frac{1}{\lambda_{st}} \left(p_{st}^\mu - \frac{y_{(st)Q}}2 Q^\mu\right)
+ \frac{y_{(st)Q}-y_{st}}{2} Q^\mu,\\
& \tilde p_i^\mu = \Lambda^\mu{}_\nu\left[Q-\tilde p_{st}, Q-p_{st}\right]\, p_i^\nu \quad\text{for}\quad i\neq s,t,\\
& \text{where}\quad \lambda_{st}= \frac{\sqrt{y_{(st)Q}^2-4y_{st}^2}}{y_{(st)Q}-y_{st}},
\end{aligned}
\label{eq:lorentzmapping:mappingtwo}
\end{equation}
and $\Lambda\left[Q-\tilde p_{st}, Q-p_{st}\right]$ is a Lorentz transform that maps the 4-vector $Q-p_{st}$ to $Q-\tilde p_{st}$, ensuring momentum conservation.%
\footnote{
Although this is not apparent in \cref{eq:lorentzmapping:mappingtwo}, the square of $\lambda_{st}$ is a ratio of K\"allen functions, as one can see from \cref{eq:rescpar}.}
It is possible to define such a transformation independently of the space-time dimension as
\begin{equation}
	\Lambda^\mu{}_\nu[\tilde{K}, K] = g^\mu{}_\nu
- \frac{2(K+\tilde{K})^\mu(K+\tilde{K})_\nu}{(K+\tilde{K})^{2}} 
+ \frac{2\tilde{K}^\mu K_\nu}{K^2}.
\label{eq:lorentzlambda}
\end{equation}
Such a Lorentz transformation only exists if the two momenta have the same non-vanishing invariant mass $K^2 = \tilde{K}^2$.
The expression is therefore valid unless there is only one massless final-state recoiler.
When this situation arises, as e.g.\ in the case of dijet production, another form of the Lorentz transformation $\Lambda$ must be adopted, as for instance the one described in \cref{app:lorentztransform}.
It is easy to verify that indeed \cref{eq:lorentzlambda} gives $\Lambda[\tilde{K}, K]  K = \tilde K$ and $\Lambda \cdot \Lambda^T = \mathbb{1}$.

The mapping \labelcref{eq:lorentzmapping:mappingtwo} is again generalisable to $k$ unresolved massless momenta $p_{t_1},\dots, p_{t_k}$,
\begin{equation}
\begin{aligned}
&  \tilde p_{t_1\dots t_k}^\mu =  \frac{1}{\lambda_{t_1\dots t_k}}\, \left( p_{t_1\dots t_k}^\mu - \frac{y_{(t_1\dots t_k)Q}}2 Q^{\mu}\right)
+ \frac{y_{(t_1\dots t_k)Q}-y_{t_1\dots t_k}}{2} Q^{\mu},\\
&  \tilde p_i = \Lambda\left[Q-\tilde p_{t_1\dots t_k}, Q-p_{t_1\dots t_k}\right]\, p_i \quad\text{for}\quad i\neq t_1,\dots, t_k,\\
&\text{where}\quad \lambda_{t_1\dots t_k}= \frac{\sqrt{y_{(t_1\dots t_k)Q}^2-4y_{t_1\dots t_k}^2}}{y_{(t_1\dots t_k)Q}-y_{t_1\dots t_k}}.
\end{aligned}
\end{equation}
The phase space factorisation for $n = m+k-1$ final-state particles is
\begin{align}
\int \dd{\Phi_{n}(\{p\}_{n};Q)} &= \int \dd{\Phi_m(\{\tilde p\}_m;Q)}
\int_0^{s_0}\frac{\dd{s_{t_1\dots t_k}}}{2\pi} \lambda^{d-3} \int 
\dd{\Phi_k\left(\left\{p_{t_i}\right\}_k;p_{t_1\dots t_k}\right)},
\label{eq:lorentzmapfactorisation}
\end{align}
where $s_0=Q^2 \left(1-\sqrt{1-\tilde y}\right)^2$.

Note that, as shown in \cref{app:massive_lorentz}, this mapping can be generalised to the case where some of the unresolved momenta, as well as the parent momentum $\tilde p_{t_1\dots t_k}$, are massive, which makes it suitable for quasi-collinear counterterms in processes with massive coloured particles.

\subsection{Initial-collinear mappings}

Initial-collinear mappings are used to subtract divergences that occur when a set of final-state particles become collinear to an initial-state particle.
There are two main differences compared to final-collinear mappings,
\begin{itemize}
	\item initial-collinear mappings generate a convolution on Bjorken momentum fractions that cannot be cast into a factorised form;
	\item the unresolved phase space contains one particle less than the number of final-state particles that are not resolved.
\end{itemize}
In this section, we provide a single example of mapping, which was used at \gls{NLO} in the original dipole subtraction~\cite{Catani:1996vz}, 
in antenna subtraction~\cite{Daleo:2006xa}, in \colorful\ subtraction~\cite{Somogyi:2009ri} as well as in \gls{FKS} subtraction~\cite{Frixione:1995ms}.
Let us consider the factorisation of the amplitude for a process $p_a,p_b\to p_1,\dots, p_{m+1}$ when $p_{m+1}\parallel p_a$,
\begin{align}
\lim_{p_{m+1}\parallel p_a}\left|{\cal M}\left(p_1,\dots,p_{m+1};p_a,p_b\right)\right|^2 = 8\pi \alpha_s \mu^{2\epsilon}\frac{P(x)}{x\,s_{a,m+1}} \left| {\cal M}\left(p_1,\dots,p_{m};p_a-p_{m+1},p_b\right)\right|,
\end{align}
where
\begin{align}
x = \frac{(p_{a}-p_{m+1})\cdot n}{p_a\cdot n},
\label{eq:initialcoll:singlexdef}
\end{align}
for any reference 4-vector $n$ such that $n\cdot p_a\ne 0$.
We aim to parametrise the $(m+1)$-particle phase space with total momentum $Q=p_a+p_b$ in terms of variables which describe the $(m+1)$-th emission and an $m$-particle phase space with total momentum $\tilde{Q}=x p_a + p_b$.
Contrary to the case of final-collinear mappings, we thus not only have to change the final-state momenta but also the initial-state ones.
We set $n=Q$ and realise the mapping as follows,
\begin{align}
\tilde p_a^\mu & = \frac{(p_a-p_{m+1})\cdot Q}{p_a\cdot Q} p_a^\mu = x p_a^\mu, \qquad
\tilde p_b^\mu = p_b^\mu, \nonumber\\
\tilde p_i^\mu &= \Lambda^\mu{}_\nu\left[\tilde{Q},Q-p_{m+1}\right] p_i^\nu \quad\text{for}\quad 1 \leq i \leq m. 
\end{align}
It is easy to observe that this mapping does indeed achieve its intended goal: in the $p_{m+1}\parallel p_a$ limit, $\tilde p_a = p_a - p_{m+1}$, so $\tilde Q= Q-p_{m+1}$ and therefore also $\tilde p_i = p_i$.
Note that all final-state particles need to be shifted for the mapping to work.
The phase space can be reparametrised in terms of the new momenta,
\begin{align}
&\int \dd{\Phi_{n}(\{p\}_{n};p_a+p_b)} \nonumber\\
&\qquad = \int_0^1 \dd{\xi} \int \dd{\Phi_{m}(\{\tilde p\}_{m};\xi p_a+p_b)} \int \frac{\dd[d]{p_{m+1}}}{(2\pi)^{d-1}}\delta_+\brk*{p_{m+1}^2} \delta(\xi-x)
\nonumber\\
&\qquad =\int_0^1 \dd{\xi} \int \dd{\Phi_{m}(\{\tilde p\}_{m};\xi p_a+p_b)} \int \brk[s]*{\dd{p_{m+1}}}(\xi),
\end{align}
where $x$ in the second line is given by \cref{eq:initialcoll:singlexdef} and is therefore a function of $p_a$, $p_{m+1}$ and $n = Q$.
Using this mapping we do not completely factorise the phase space, but obtain a convolution where $\xi$ entangles the energy of the emitted unresolved particle with the resolved initial state momentum.
As a result, counterterms integrated over the unresolved degrees of freedom $[\dd{p_{m+1}}]$ will feature a dependence on $\xi$.

As in the case of final-collinear mappings, this transformation can be generalised without effort to $k$ particles becoming collinear to the initial momentum $p_a$~\cite{Daleo:2006xa},
\begin{align}
\tilde p_a^\mu & = \frac{\left(p_a-p_{m+1}-\dots- p_{m+k}\right)\cdot Q}{p_a\cdot Q} p_a^\mu = x p_a^\mu, \qquad
\tilde p_b^\mu = p_b^\mu, \nonumber\\
\tilde p_i^\mu &= \Lambda^\mu{}_\nu\brk[s]*{\tilde{Q}, Q - p_{m+1} - \dots - p_{m+k}} p_i^\nu \quad\text{for}\quad 1 \leq i \leq m,
\end{align}
yielding the phase space convolution,
\begin{align}
& \int \dd{\Phi_{n}(\{p\}_{n};p_a+p_b)} \nonumber\\
&\qquad = \int_0^1 d \xi \int \dd{\Phi_{m}(\{\tilde p\}_{m};\xi p_a+p_b)}
\int \brk[s]*{\prod_{i=1}^k \frac{\dd[d]{p_{m+i}}}{(2\pi)^{d-1}}\delta_+\brk*{p_{m+i}^2}} \delta(\xi-x) \nonumber\\
& \qquad =\int_0^1 d \xi \int \dd{\Phi_{m}(\{\tilde p\}_{m};\xi p_a+p_b)} \int \brk[s]*{\dd{p_{m+1}}\dots\dd{p_{m+k}}}(\xi).
\end{align}

\subsection{Soft mappings for massless partons}
\label{sec:softmappings}
\input{softmappingreview}

%% file: softmappingreview.tex
The soft mapping of \cref{sec:softexample} is trivially extended to $m$ massless partons of momenta $p_1,\ldots,p_m$
and one soft gluon of momentum $p_s$.
The $m$ mapped momenta are defined by first rescaling all the hard momenta by a factor $1/\lambda$ and then Lorentz-transforming all of the rescaled momenta \cite{Somogyi:2006cz}
\begin{equation}
\tilde{p}_i^\mu = \Lambda^\mu{}_\nu\brk[s]*{Q,\frac{Q-p_s}{\lambda}} \frac{p_i^\nu}{\lambda} \quad\text{for}\quad 1\le i\le m.
\label{eq:PS_Sr}
\end{equation}
Here $\Lambda$ is given in \cref{eq:lorentzlambda} with $\tilde{K} = Q$ and $K = (Q-p_s)/\lambda$.
The constraint $K^2 = \tilde{K}^2$ implies that
\begin{equation}
\lambda = \sqrt{1-y_{sQ}}.
\label{eq:lambdar}
\end{equation}
The phase space of \cref{eq:softphasespace} is generalised to $m$ hard partons,
\beq
\PS[m+1](\{p\}_{m+1};Q)=\PS[m](\{\tilde{p}\}_m; Q) \lambda^{(m-1)(d-2)-2}
\frac{\dd[d]{p_s}}{(2\pi)^{d-1}} \delta_+\brk*{p_s^2}\theta(\lambda) \theta(1-\lambda),
\label{eq:PSfact_Sr}
\eeq
where the $m$ momenta in the first factor on the \acrlong{RHS} are those of \cref{eq:PS_Sr}.

It is straightforward to extend the single-soft mapping of \cref{eq:PS_Sr} to a multiple-soft mapping for $m$ massless hard partons of momenta $p_1,\ldots,p_m$ and $r$ soft partons of momenta $p_{s_1},\ldots, p_{s_r}$, with $n=m+r$.
The $m$ momenta $\ti{p}_1,\ldots,\ti{p}_m$ are given by
\beq
\ti{p}_i^\mu =
\Lambda^\mu{}_\nu\brk[s]3{Q,\frac{Q-\sum_{j=1}^r p_{s_j}}{\lambda_{s_1\ldots s_r}}} \frac{p_i^\nu}{\lambda_{s_1\ldots s_r}} \quad\text{for}\quad 1\le i\le m,
\label{eq:PS_Smanys}
\eeq
where $\Lambda$ is given by \cref{eq:lorentzlambda} with $\ti{K}=Q$ and $K=(Q-\sum_{j=1}^r p_{s_j})/\lambda_{s_1\ldots s_r}$. 
The constraint $K^2 = \tilde{K}^2$ implies that
\beq
\lambda_{s_1\ldots s_r} = \sqrt{1-\left(y_{(s_1\ldots s_r)Q} - y_{s_1\ldots s_r}\right)}.
\label{eq:lambdamanys}
\eeq
The phase space of \cref{eq:PSfact_Sr} is generalised to the mapping of \cref{eq:PS_Smanys},
\begin{align}
&\PS[n](\{p\}_{n};Q) \nn\\
&= \PS[m](\{\tilde{p}\}_m; Q) \lambda_{s_1\ldots s_r}^{(m-1)(d-2)-2} \brk[s]3{\prod_{j=1}^r \frac{\dd[d]{p_{s_j}}}{(2\pi)^{d-1}} \delta_+(p_{s_j}^2)}
\theta(\lambda_{s_1\ldots s_r}) \theta(1-\lambda_{s_1\ldots s_r}).
\label{eq:PSfact_Smanys}
\end{align}
In the case of two soft partons, the mapping of \cref{eq:PS_Smanys} was used in \cite{Somogyi:2006da}.

%% file: finalcoll.tex
\section{Generalised rescaling mapping}
\label{sec:fcoll}

In this section, we introduce a transformation which reparametrises the $m$-particle phase space of the momenta $\{p\}_m$ with masses $\{m\}_m$ in terms of $m$ momenta $\{\tilde p\}_m$ with different masses $\{\tilde m\}_m$ and the same total momentum $Q$.
We propose a novel application of this transformation as a final-collinear momentum mapping.
The main application is of course the subtraction of genuine \gls{IRC} singularities, where it is used to replace momenta of sets of massless final-state particles going collinear to each other with on-shell momenta for their massless parents.
We foresee that other applications will be relevant, such as the stabilisation of quasi-collinear singularities in processes with massive coloured particles.

We introduce the transformation in \cref{ssec:genresc:def}, then we outline its usage as a mapping and derive the corresponding phase-space factorisation in \cref{ssec:genresc:mapping}.
We highlight important properties of this mapping in \cref{ssec:genresc:commass}, give an explicit application in \cref{ssec:genresc:appl} and comment on counterterm integration in \cref{ssec:genresc:jac}.
Finally, in \cref{ssec:genresc:cases} we point out several special cases in which the transformation is significantly simpler to formulate or reduces to one of the mappings defined in the previous section.

\subsection{Definition}
\label{ssec:genresc:def}

We begin by defining how the mapping acts on the $\{p\}_m$ phase space.
Although the mapping can be formulated in a manifestly covariant form, as we shall see in \cref{eq:explLorentz},
for the sake of simplicity we work in the rest frame of the total momentum $Q$ and use non-explicitly Lorentz-covariant notation.
In the considered frame, the 3-momenta involved in the mapping sum to zero,
\begin{equation}
	\sum_i \vect{p}_i = \vect{0}.
\end{equation}
Spatial momentum conservation therefore remains valid if all 3-momenta are rescaled by a common arbitrary factor $\resc$,
\begin{equation}
\label{eq:fcoll:spaceresc}
	\vect{\tilde{p}}_i = \frac{1}{\resc} \vect{p}_i.
\end{equation}
Energies can then be set by imposing the mass-shell conditions,
\begin{equation}
	\tilde{E}_i = \sqrt{\vect{\tilde{p}}^2_i + \tilde{m}^2_i},
\end{equation}
and the parameter $\resc$ finally be fixed by requiring energy conservation,
\begin{equation}
\label{eq:fcoll:eqforresc}
	\sum_i \tilde{E}_i = \sum_i E_i,
	\qquad\text{i.e.}\qquad
	\sum_i \sqrt{\vect{p}^2_i/\resc^2 + \tilde{m}^2_i} = Q,
\end{equation}
where we abuse notation by using $Q$ to refer to $\sqrt{Q^2}$.
Once the values for the target masses $\{\tilde{m}\}_m$ are given, this is an algebraic equation for the unknown $\resc$.
The \acrlong{LHS} of \cref{eq:fcoll:eqforresc} is a monotonous function of $\resc$ which varies between $\sum_i \tilde m_i$ and $\infty$ over $\resc \in \mathbb{R}^+$, and therefore admits a unique valid solution as long as the physical condition $\sum_i \tilde m_i\leq Q$ is respected.%
\footnote{Note that if $\tilde{m}_i \le m_i$ for all particles $i$, the actual range for the solution is $0<\resc\le 1$.}
The solution for $\resc$ can be promptly written in closed form if only two or three momenta are involved.
For a generic number $m$ of momenta $\{p\}_m$, one can also show that any solution of \cref{eq:fcoll:eqforresc} is one of the solutions of a polynomial equation%
\footnote{This polynomial can be obtained constructively by isolating one square root and squaring the equation.}
of degree $2^{2m-1}$, indicating that the general case requires a numerical solution.
In any case, it is immediate to see that if the target masses do not change, $\tilde{m}_i = m_i$, one finds $\resc = 1$ and the mapping reduces to the identity.

We have so far provided a procedure to generate $m$ momenta with masses $\{\tilde m\}_m$ from $m$ momenta with masses $\{m\}_m$ such that their total momentum $Q$ is left unchanged.
Let us now show how this affects the phase-space measure.
For now we formulate the problem as a change of variables in the $m$-particle phase space integration, and we will cast this general approach to specific cases of phase-space factorisations for subtraction in the next section.

Working in the rest frame of $Q$,
the original $m$-particle phase space reads
\begin{equation}
	\dd{\Phi}(Q^2; \set{m}_m) =
	(2\pi)^d \delta\brk3{Q - \sum_i E_i}
	\delta^{(d-1)}\brk3{\sum_i \vect{p}_i}
	\brk[s]3{
		\prod_i \frac{1}{2E_i} \frac{\dd[d-1]{\vect{p}_i}}{(2\pi)^{d-1}}
	},
\end{equation}
where
\begin{equation}
	E_i = \sqrt{\vect{p}^2_i + m^2_i}.	
\end{equation}
In order to rewrite it in terms of the mapped momenta,
it is useful to insert the identity,
\begin{equation}
	1
	= \int\dd{\resc'} \delta\brk{\resc'-\resc}
	= \int\dd{\resc'}
	\delta\brk3{Q - \sum_i \sqrt{\vect{p}^2_i/\resc'^2 + \tilde{m}^2_i}}
	\brk[s]3{\sum_i\frac{\vect{p}^2_i}{\resc^3 \tilde{E}_i}}.
\end{equation}
Changing variables according to $\vect{p}_i=\resc\vect{\tilde{p}}_i$ then gives
\begin{multline}
\label{eq:fcoll:PSabc}
	\dd{\Phi}(Q^2; \set{m}_m) =
	(2\pi)^d
	\delta\brk3{Q - \sum_i \tilde{E}_i}
	\delta^{(d-1)}\brk3{\sum_i \vect{\tilde{p}}_i}
	\brk[s]3{
		\prod_i \frac{1}{2\tilde{E}_i}
		\frac{\dd[d-1]{\vect{\tilde{p}}_i}}{(2\pi)^{d-1}}
	}
	\\\times
	\resc^{1-d}
	\brk[s]3{\prod_i \resc^{d-1}\frac{\tilde{E}_i}{E_i}}
	\dd{\resc'}
	\delta\brk3{Q - \sum_i \sqrt{\resc'^2\vect{\tilde{p}}^2_i + m^2_i}}
	\brk[s]3{\sum_i\frac{\vect{\tilde{p}}^2_i}{\resc \tilde{E}_i}},
\end{multline}
which yields
\begin{align}
	\dd{\Phi}(Q^2; \set{m}_m)&=
	\dd{\tilde{\Phi}}(Q^2; \set{\tilde{m}}_m)
	\times \resc^{-d-1}
	\brk[s]3{\prod_i \resc^{d-1} \frac{\tilde{E}_i}{E_i}}
	\brk[s]3{\sum_i\frac{\vect{\tilde{p}}^2_i}{E_i}}^{-1}
	\brk[s]3{\sum_i\frac{\vect{\tilde{p}}^2_i}{\tilde{E}_i}} \nonumber\\
	&=\dd{\tilde{\Phi}}(Q^2; \set{\tilde{m}}_m)
	\times J\left(\set{\tilde{p}}_m,\set{\tilde{m}}_m,Q\right), \label{eq:fcoll:genrescjac}
\end{align}
where we introduced the Jacobian of the transformation $J$.
We have checked this result by numerically computing the phase-space volume obtained by integrating over the two phase-space parametrisations for arbitrary choices of masses and up to $m=7$ particles.

This result is formulated in terms of non-manifestly covariant quantities, but since we worked in the rest frame of $Q$ we can write
\begin{align}
E_i &= \frac{p_i\cdot Q}{\sqrt{Q^2}}, &
\vect{p}^2_i &=m^2_i - \frac{(p_i\cdot Q)^2}{Q^2},
\label{eq:explLorentz}
\end{align}
in order to restore explicit Lorentz covariance.

\subsection{Mapping of multiple clusters of collinear particles}
\label{ssec:genresc:mapping}

Let us now see how the transformation applies to clusters of particles becoming collinear to each other.
For a single cluster of $k$ massless momenta within an $n$-particle phase space,
\begin{align}
\int \dd{\Phi_{n}(\{p\}_n;Q)}
&= \int_{0}^{s_0} \frac{\dd{s_K}}{2\pi}\int \dd{\Phi_{m}(\set{(p_i,m_i)}_{m-1},(p_K,\sqrt{s_K});Q)}\int \dd{\Phi_{k}(\set{p_i}_{k};Q)} \nonumber\\
& = \int \dd{\Phi_{m}(\left\{(\tilde p_i,m_i)\right\}_{m};Q)} \int_{0}^{s_0} \frac{\dd{s_K}}{2\pi} J(\left\{\tilde p_i\right\}_{m},s_K,Q) \int \dd{\Phi_{k}(\left\{p_i\right\}_{k};Q)}\,.
\end{align}

\input{multi_collinear_map_fig}

By virtue of our ability to map several massive momenta to massless momenta and the easy generalisation of \cref{eq:decayfactorisation} to several splittings, we can provide a phase-space factorisation that suits the subtraction of $N$ clusters of $k_1,\dots,k_N$ particles becoming collinear as depicted in \cref{fig:multicol}.
The expression is as follows,
\begin{align}
\int& \dd{\Phi_{n}(\{p\}_{n};Q)}=\nonumber\\
&\phantom{=}\int_{0}^{s_0^{K_1}} \frac{\dd{s_{K_1}}}{2\pi}\times\dots\times\int_{0}^{s_0^{K_N}} \frac{\dd{s_{K_N}}}{2\pi}\int \dd{\Phi_{m}\left(\left\{\left(p_i,m_i\right)\right\}_{m-N},\left\{\left(p_{K_j},\sqrt{s_{K_j}}\right)\right\}_N;Q\right)}\nonumber\\
&\times\int \dd{\Phi_{k_1}(\left\{p_i\right\}_{k_1};p_{K_1})}\times\dots\times \int \dd{\Phi_{k_N}(\left\{p_i\right\}_{k_N};p_{K_N})} \nonumber\\
&=\int \dd{\tilde\Phi_{m}\left(\left\{\left(\tilde p_i,m_i\right)\right\}_{m};Q\right)}\int_{0}^{s_0^{K_1}} \frac{\dd{s_{K_1}}}{2\pi}\times\dots\times\int_{0}^{s_0^{K_N}} \frac{\dd{s_{K_N}}}{2\pi} J\left(\left\{\tilde p_i\right\}_{m}\right),\left\{s_K\right\}_N,Q)\nonumber \\
&\times\int \dd{\Phi_{k_1}(\left\{p_i\right\}_{k_1};p_{K_1})}\times\dots\times \int \dd{\Phi_{k_N}(\left\{p_i\right\}_{k_N};p_{K_N})}\,.
\label{eq:PS_fact_genmap}
\end{align}
As a specific example, this allows us to express the $q\,\bar q\, g\, g\, g$ phase space $\Phi(p_q, p_{\bar q}, p_{g_1}, p_{g_2}, p_{g_3})$ in a factorised way of the form $\dd{\Phi(\tilde q\, \tilde{\bar q}\, \tilde g_3)} \times \dd{\Phi (q\to q\, g_1)} \times \dd{\Phi(\bar{q}\to \bar{q} \,g_2)}$ in order to define counterterms for the ``double-collinear'' limit where gluon $g_1$ becomes collinear to the quark and gluon $g_2$ becomes collinear to the antiquark.
This case arises in double-real radiative corrections to the final state $q\, \bar{q}\, g$ at \gls{NNLO}.

\subsection{Commutativity and associativity}
\label{ssec:genresc:commass}

\input{mapping_properties_fig}

Processes where multiple disjoint clusters of massless particles can become collinear at the same time feature multiple singular limits that require subtraction.
As mentioned above, one singular kinematic configuration is the one where all the children particles become collinear to their respective parents at once.
However, the limits where only one or some children clusters become collinear are also divergent and need to be subtracted, and a pattern of cancellation between the different counterterms is required for the subtraction to work.
As a result, it is useful to have a mapping that ensures that the different counterterms for these sub-limits yield reduced phase-space points that match under appropriate conditions.
This is guaranteed to happen if the properties of \emph{associativity} and \emph{commutativity} are respected by the mapping.
\begin{itemize}
	\item \textbf{Commutativity} is realised if mapping a process with multiple separate simultaneous splittings sequentially (i.e.\ splitting by splitting) yields the same reduced phase-space point independently of the order chosen, as illustrated in \cref{fig:commutativity}.
	\item \textbf{Associativity} is realised if mapping a single splitting with multiple children in one step or sequentially merging subsets of the children yields the same reduced phase-space point, as illustrated in \cref{fig:associativity}.
\end{itemize}
Let us see how these properties are realised in the case of the generalised rescaling mapping.

Commutativity is easy to prove.
Take $N$ clusters of particles whose sums of constituent momenta $\set{p_1,\dots,p_N}$ are mapped one after the other to on-shell parent momenta recoiling against $m$ other momenta $\set{q_1,\dots,q_m}$.
Let $\sigma$ be the permutation of $1,\dots,N$ that specifies the order in which clusters are merged into their parents.
Each step leaves the direction of the clusters' or parent 3-momenta unchanged and rescales them by a parameter $\kappa_i^{\sigma}$, where $i$ labels the step .
Whatever the order $\sigma$ of the sequential cluster merging, the final mapped phase space $\set{\tilde p_1,\dots,\tilde p_N,\tilde q_1,\dots,\tilde q_m}$ verifies the following energy-conservation equation,
\begin{align}
\sum_{i=1}^N \sqrt{\frac{1}{\kappa_1^\sigma\dots \kappa_N^\sigma}\vec{p}_i^{\,2}+\tilde m^2_i} + \sum_{j=1}^m \sqrt{\frac{1}{\kappa_1^\sigma\dots \kappa_N^\sigma}\vec{q}_j^{\,2}+\tilde m_j^2} = Q.
\end{align}
As we already argued, the left hand side is a monotonous function of $\kappa^\sigma=\kappa_1^\sigma\dots \kappa_N^\sigma$ over $\mathbb{R}^+$, so that there is a unique physical solution for the final phase-space point independently of the order of the iterated mappings $\sigma$.
By the same argument, the result is also independent of whether multiple clusters are merged simultaneously or one after the other.

The proof for associativity follows along the same line.
Without loss of generality, let us consider for simplicity the case of three momenta $\set{p_1,p_2,p_3}$ with masses $\set{m_1,m_2,m_3}$ being combined into a momentum $\tilde p_{123}$ with mass $m_{123}$.
If we perform the mapping in one step, the vector direction $\vec p_{123}$ is kept unchanged and we need to solve for $\kappa_{123}$ in
\begin{align}
\sqrt{\frac{1}{\kappa_{123}} \vec{p}_{123}^{\,2} + m_{123}^2}+ \sum_{j\ne1,2,3} \sqrt{\frac{1}{\kappa_{123}}\vec{q}_j^{\,2}+ m_j^2} = Q.
\end{align}
If we first map the momenta $p_2,p_3$ into an intermediate momentum $\hat{p}_{23}$ with mass $m_{23}$, and then map $\hat{p}_{23}$ with $\hat{p}_{1}$, we have
\begin{gather}
\vect{\hat{p}}_{23} = \frac{1}{\kappa_{23}} (\vect{p}_2+\vect{p}_3), \qquad
\vec{\hat{p}}_1 = \frac{1}{\kappa_{23}} \vec{p}_1, \qquad
\vec{\hat{q}}_j = \frac{1}{\kappa_{23}} \vec{q}_j, \\
\vec{\bar{\hat{p}}}_{123} = \frac{1}{\kappa_{1,23}}(\vec{\hat{p}}_1+\vec{\hat{p}}_{23}) = \frac{1}{\kappa_{1,23}\kappa_{23}} \vec{p}_{123}, \qquad
\vec{\bar{\hat{q}}}_j = \frac{1}{\kappa_{1,23}\kappa_{23}} \vec{q}_j,
\end{gather}
where $\kappa_{1,23}\kappa_{23}$ must verify the same energy conservation condition as $\kappa_{123}$ and the final parent momentum spatial direction is still that of $p_{123}$.
As a result, $\vect{\bar{\hat{p}}}_{123} = \vect{\tilde{p}}_{123}$.
Furthermore, any other momentum $q_j$ in the process is mapped by having its spatial components rescaled, yielding the same final momentum as well.
The result of the mapping is thus independent of whether the particles of a cluster are merged all at once or one after the other.

\subsection{Application to $e^+\,e^- \to q\,\bar{q}\,q'\,\bar{q}'\,g$}
\label{ssec:genresc:appl}

Associativity and commutativity make the subtraction of iterated limits in schemes without sectors considerably more straightforward.
Let us illustrate this with the example of a double-unresolved limit of $e^+\,e^- \to q\,\bar{q}\,q'\,\bar{q}'\,g$, where $q$ and $q'$ are quarks of different flavour, and $\bar{q}$ and $\bar{q}'$ are their respective antiquarks.
We number the particles as $q_1,\,\bar{q}_2,\,q'_3,\,\bar{q}'_4,\,g_5$.

The squared amplitude $\abs{{\cal M}}^2$ features, amongst others, two singularities in the limits where either quark-antiquark pair becomes collinear and the other has generic kinematics, $C_{12}$ and $C_{34}$.
These divergences need to be regulated if integration is to be performed in 4 space-time dimensions.
As discussed above, this can be achieved using the factorisation properties of the squared amplitude and a momentum mapping to build counterterms as follows,
\begin{align}
\begin{cases}
	P_{12}\abs{{\cal M}}^2\left(\tilde p_{12},\tilde p_3,\tilde p_4, \tilde p_5\right) \text{ approximates } \abs{{\cal M}}^2\left(p_1,p_2,p_3,p_4,p_5\right) \text{ in }C_{12},\\
	P_{34}\abs{{\cal M}}^2\left(\hat p_1,\hat p_2,\hat p_{34}, \hat p_5\right) \text{ approximates } \abs{{\cal M}}^2\left(p_1,p_2,p_3,p_4,p_5\right) \text{ in }C_{34},
\end{cases}
\label{eq:noncommsinglects}
\end{align}
where $P_{ij}$ is the splitting kernel for the appropriate limit and we omit spin-correlation indices.
Tilded and hatted momenta indicate the mapped momenta in the mappings for the limits $C_{12}$ and $C_{34}$ respectively.

In the limit $C_{12,34}$ where both quark-antiquark pairs are collinear to each other, the counterterms designed for the collinear configurations $C_{12}$ and $C_{34}$ both approximate the matrix element thus leading to over-subtraction.
Moreover, each of these two counterterms for singular single-unresolved configurations in turn presents a divergence in the region of phase space where the opposite mapped quark-antiquark pair goes collinear, i.e.\ $\tilde{p}_3\parallel\tilde{p}_4$ and $\hat{p}_1\parallel\hat{p}_2$.
We denote the limits which approach these singular kinematics with $C_{\tilde{3}\tilde{4}}$ and $C_{\hat{1}\hat{2}}$.
Note that in general the loci of the limits $C_{\tilde{3}\tilde{4}}$ and $C_{\hat{1}\hat{2}}$ and those of $C_{34}$ and $C_{12}$ do \emph{not} coincide.
In order to regulate the divergences associated to these double-unresolved configurations, a counterterm for the limit $C_{12,34}$ and counter-counterterms for the limits $C_{\hat{1}\hat{2}}C_{34}$ and $C_{\tilde{3}\tilde{4}}C_{12}$ need to be introduced,
\begin{align}
\begin{cases}
	P_{12}P_{34}\abs{{\cal M}}^2(\bar{p}_{12},\bar{p}_{34},\bar{p}_5)
	\text{ approximates }
	\abs*{{\cal M}}^2\brk*{p_1,p_2,p_3,p_4,p_5}
	\text{ in }C_{12,34},
	\\
	P_{12}P_{\tilde{3}\tilde{4}}\abs{{\cal M}}^2
	(\hat{\tilde{p}}_{12},\hat{\tilde{p}}_{34},\hat{\tilde{p}}_5)
	\text{ approximates }
	P_{12}\abs{{\cal M}}^2(\tilde{p}_{12},\tilde{p}_3,\tilde{p}_4,\tilde{p}_5)
	\text{ in }C_{\tilde{3}\tilde{4}},
	\\
	P_{\hat{1}\hat{2}}P_{34}\abs{{\cal M}}^2
	(\tilde{\hat{p}}_{12},\tilde{\hat{p}}_{34},\tilde{\hat{p}}_5)
	\text{ approximates }
	P_{34}\abs{{\cal M}}^2(\hat{p}_1,\hat{p}_2,\hat{p}_{34},\hat{p}_5)
	\text{ in }C_{\hat{1}\hat{2}}.
\end{cases}
\label{eq:noncommdoublects}
\end{align}

Starting from the singularities which correspond to the two single-unresolved configurations $C_{12}$ and $C_{34}$, we obtained an integrand which contains six terms: the original squared matrix element and the five counterterms of \cref{eq:noncommsinglects,eq:noncommdoublects}.
This situation is illustrated in \cref{fig:nocommut}.
In the limit $C_{12,34}$ both quark-antiquark pairs are collinear, and as a consequence $\tilde{p}_{3/4}\to p_{3/4}$ and $\hat{p}_{1,2}\to p_{1/2}$.
Therefore we also asymptotically have $\tilde{p}_3\parallel\tilde{p}_4$ and $\hat{p}_1\parallel\hat{p}_2$, all mappings reduce to the identity and in the exact limit we have
\begin{equation}
	C_{12,34}:\qquad
	\bar{p}_{12} = \tilde{\hat{p}}_{12} = \hat{\tilde{p}}_{12},
	\quad
	\bar{p}_{34} = \tilde{\hat{p}}_{34} = \hat{\tilde{p}}_{34}.
\end{equation}
The three terms in \cref{eq:noncommdoublects} have matrix elements that are evaluated for the same phase-space point which makes it possible for simple cancellation patterns to take place.
However in one of the single-unresolved limit, say $C_{12}$, we only find $\tilde{p}_{3/4}\to p_{3/4}$ but in general $\hat{p}_{1,2}\ne p_{1/2}$.
The mapping $C_{34}$ need not reduce to the identity and neither does $C_{\tilde{3}\tilde{4}}$.
This means that the three terms in \cref{eq:noncommdoublects} contain matrix elements (and in general measurement functions) that are evaluated at \emph{different} phase-space points and it is highly non-trivial for cancellations to occur.
This observation alone does not exclude that there might be a way to subtract all overlaps with a clever choice of counterterms, but it is clear that non-commutativity makes engineering iterative counter-counterterms a highly non-trivial task.

A much simpler situation can be achieved using a commutative mapping.
Indeed, commutativity ensures that the iterated counter-counterterms for the $C_{12}$ and $C_{34}$ limits have the same reduced kinematics as the $C_{12,34}$ counterterm for \emph{any} phase-space point, as illustrated in \cref{fig:yescommut}.
This is for example exploited in \colorful\ subtraction~\cite{Somogyi:2005xz,Somogyi:2006da,Somogyi:2006db,Somogyi:2008fc,Aglietti:2008fe,Somogyi:2009ri,Bolzoni:2009ye,Bolzoni:2010bt,DelDuca:2013kw,Somogyi:2013yk},
antenna subtraction~\cite{GehrmannDeRidder:2005cm,Daleo:2006xa,GehrmannDeRidder:2005aw,GehrmannDeRidder:2005hi,Daleo:2009yj,Gehrmann:2011wi,Boughezal:2010mc,GehrmannDeRidder:2012ja,Currie:2013vh}
and local analytic sector subtraction~\cite{Magnea:2018hab,Magnea:2018ebr}.
On the other hand, the Lorentz mapping, which one could hope to use as an alternative for the rescaling mapping of \colorful\ for massive final states, is neither commutative nor associative when there are more than two particles in the Born final state.

\begin{figure}
\begin{subfigure}{\textwidth}
	\centering
	\begin{tikzcd}[row sep=huge,column sep=large]
	\abs{{\cal M}}^2(p_1,p_2,p_3,p_4,p_5)\arrow[rd, "{C_{12,34}}"]\arrow[r, "C_{12}"]\arrow[d, "C_{34}"] & P_{12} \abs{{\cal M}}^2 (\tilde p_{12},\tilde p_3,\tilde p_4, \tilde p_5)\arrow[rd,"C_{\tilde 3 \tilde 4}"] &
	\\P_{34}\abs{{\cal M}}^2(\hat p_1,\hat p_2,\hat p_{34},\hat p_5)\arrow[rd,"C_{\hat 1 \hat 2}"]& P_{12}P_{34}\abs{{\cal M}}^2(\bar p_{12},\bar p_{34}, \bar p_5)\arrow[r,"?",leftrightarrow]\arrow[d,"?",leftrightarrow]&P_{12}P_{\tilde 3\tilde 4}\abs{{\cal M}}^2(\hat{\tilde{p}}_{12},\hat{\tilde{p}}_{34}, \hat{\tilde{p}}_5)\\
	{}&P_{\hat 1 \hat 2}P_{34}\abs{{\cal M}}^2(\tilde{\hat{p}}_{12},\tilde{\hat{p}}_{34}, \tilde{\hat{p}}_5)&{}
	\end{tikzcd}
	\caption{No commutativity}
	\label{fig:nocommut}
\end{subfigure}
\\[1em]
\begin{subfigure}{\textwidth}
	\centering
	\begin{tikzcd}[row sep=5.5em,column sep=huge]
	\abs{{\cal M}}^2(p_1,p_2,p_3,p_4,p_5)\arrow[rd, "{C_{12,34}}"]\arrow[r, "C_{12}"]\arrow[d, "C_{34}"] & P_{12} \abs{{\cal M}}^2(\tilde p_{12},\tilde p_3,\tilde p_4, \tilde p_5) \arrow[d,"C_{\tilde 3 \tilde 4}"]
	\\P_{34}\abs{{\cal M}}^2(\hat p_1,\hat p_2,\hat p_{34}, \hat p_5)\arrow[r,"C_{\hat 1 \hat 2}"]& \begin{array}{c}
	P_{12}P_{34}\abs{{\cal M}}^2(\bar p_{12},\bar p_{34}, \bar p_5)\\
	P_{12}P_{\tilde 3\tilde 4}\abs{{\cal M}}^2(\hat{\tilde{p}}_{12},\hat{\tilde{p}}_{34}, \hat{\tilde{p}}_5)\\
	P_{\hat 1 \hat 2}P_{34}\abs{{\cal M}}^2(\tilde{\hat{p}}_{12},\tilde{\hat{p}}_{34}, \tilde{\hat{p}}_5)
	\end{array}
	\end{tikzcd}
	\caption{With commutativity}
	\label{fig:yescommut}	
\end{subfigure}
	\centering
	\caption{Comparison of the cancellation patterns of counterterms for the double collinear limit $C_{12,34}$ and its collinear sub-limits in the case of a commuting and a not commuting mapping.}
\end{figure}

\subsection{Jacobians}
\label{ssec:genresc:jac}

The phase-space factorisation of \cref{eq:PS_fact_genmap} calls for a discussion of two potential challenges related to the Jacobian of the mapping $J$:
\begin{itemize}
\item $J$ is a process-dependent function of the phase space.
	In fact, while for some other mappings such as the ones used in dipole subtraction $J$ is only a function of a fixed number of momenta, here $J$ is a function of all the momenta in the process, meaning that integrated counterterms would be process dependent.
\item The Jacobian can be obtained as an explicit function of kinematics only for simple final states, since the degree of the equation that yields $\kappa$ increases with multiplicity.
	This is an issue for analytic integration over $s_K$, which would require knowing the full dependence of $\kappa$ on this variable.
\end{itemize}
The first issue is not noted here for the first time: it was already observed in the case of the rescaling mapping used in \colorful\ subtraction for final-collinear limits.
In that case, the Jacobian features an exponent which depends on the multiplicity, as can be seen in \cref{eq:rescalingmappingPS}.
A simple but efficient solution was already proposed in Ref.~\cite{Somogyi:2008fc} and exploits the fact that the Jacobian has to reduce to $1$ in the corresponding collinear limit.
As a result any working counterterm for that limit can be divided by $J$ without spoiling the subtraction, since it yields the same leading behaviour.
The same solution can be used for this mapping, and also solves the second issue raised above.
We will come back to this point in \cref{sec:independence}, where the implications of dividing the counterterms by the Jacobian are unfolded further.

\subsection{Special cases and relations to other mappings}
\label{ssec:genresc:cases}

\subsubsection{Rescaling mapping}
One can show that the generalised rescaling mapping reduces to the rescaling mapping presented in \cref{sec:rescalingmapping} when all mapped momenta are massless, i.e.\ $\tilde{m}_i = 0$ for all $i$.
In that case, the rescaling parameter $\resc$ of the generalised rescaling mapping is given in closed form by
\begin{equation}
	\resc
	= \sum_{i} \frac{\abs{\vect{p}_i}}{Q}
	= 1 - \sum_{i} \alpha_i,
	\label{eq:rescalingmappingkappa}
\end{equation}
where we have defined
\begin{equation}
	\alpha_i
	\equiv \frac{E_i - \abs{\vect{p}_i}}{Q}
	= \frac{Q\cdot p_i}{Q^2}
	- \sqrt{\brk3{\frac{Q\cdot p_i}{Q^2}}^2-\frac{p^2_i}{Q^2}}.
\end{equation}
Note that $\alpha_i$ is essentially the scalar product of $p_i$
with the unit light-like vector $n = (1, \hat{p}_i)$ and that it is zero for all momenta with $m_i=0$, so that the sum on the right-hand side of \cref{eq:rescalingmappingkappa} effectively runs over the massive parents that are mapped to massless momenta.
The complete mapping then reads
\begin{equation}
	\begin{cases}
	\displaystyle
	\tilde{p}_i^\mu = \frac{1}{\resc}
	\brk{p_i^\mu - \alpha_i Q^\mu}
	&\text{for massive parents},\\[0.5em]
	\displaystyle
	\tilde{p}_i^\mu
	= \frac{1}{\resc} p_i^\mu
	&\text{for massless recoilers}.
	\end{cases}
\end{equation}
Under the same assumptions the Jacobian \labelcref{eq:fcoll:genrescjac} collapses to
\begin{equation}
	\dd{\Phi}(Q^2; \set{m^2_i}) =
	\dd{\tilde{\Phi}}(Q^2; \set{0})
	\times \resc^{-d}
	\brk[s]3{\prod_i \resc^{d-1} \frac{\tilde{E}_i}{E_i}}
	\brk[s]3{\sum_i \frac{\tilde{E}^2_i}{Q E_i}}^{-1}.
\end{equation}
The expressions presented above are slightly more general than those discussed in \cref{eq:rescalingmappingkappa} as they handle the case of multiple clusters of particles becoming collinear to each other simultaneously.
It is easier to observe the correspondence with the existing literature when looking at specific examples, as discussed below.

\paragraph{Rescaling mapping for a single collinear cluster}
In the even simpler case of a single final-state collinear cluster of momenta $p_K = p_m+\dots+p_{m+k-1}$
with multiple massless recoilers $p_i$, we find
\begin{equation}
	\tilde{p}_K^\mu = \frac{p_K^\mu - \alpha_K Q^\mu}{1 - \alpha_K},
	\qquad
	\tilde{p}_i^\mu = \frac{p_i^\mu}{1 - \alpha_K} \quad\text{for}\quad 1 \le i \le m-1.
\end{equation}
This mapping was used for a collinear pair
to formulate the \colorful\ scheme at \gls{NLO} in \cite{Somogyi:2006cz},
and later to subtract triple-collinear limits
in \cite{Somogyi:2006da}.

\paragraph{Rescaling mapping for two collinear pairs}
In the case of two collinear pairs $\set{k_1,k_2}$ and $\set{l_1,l_2}$ and only massless recoiling momenta $p_i$,
the rescaling mapping reduces to
\begin{align}
	&\tilde{p}_{k_1k_2}^\mu = \frac{
		p_{k_1k_2}^\mu - \alpha_{k_1k_2} Q^\mu
	}{
		1 - \alpha_{k_1k_2} - \alpha_{l_1l_2}
	},\qquad
	\tilde{p}_{l_1l_2}^\mu = \frac{
		p_{l_1l_2}^\mu - \alpha_{l_1l_2} Q^\mu
	}{
		1 - \alpha_{k_1k_2} - \alpha_{l_1l_2}
	},\nonumber\\
	&\tilde{p}_i^\mu = \frac{p_i^\mu}{
		1 - \alpha_{k_1k_2} - \alpha_{l_1l_2}
	} \quad\text{for}\quad i\ne k_1, k_2, l_1, l_2.
\end{align}
This is the expression used to handle two final-state collinear pairs
of partons with an arbitrary number of massless recoilers $r$
in \colorful\ subtraction at \gls{NNLO} \cite{Somogyi:2006da}.

\paragraph{Rescaling mapping for one collinear set and one recoiler}
In the case of a single collinear set $K$ and a single massless recoiler $r$,
we have $Q = p_r + p_K$ and using $p_r^2 = 0$ we find
\begin{equation}
	\alpha_K = \frac{p_K^2}{Q^2},
	\qquad
	\tilde{p}_K^\mu = \frac{p_K^\mu - \alpha_K p_r^\mu}{1-\alpha_K},
	\qquad
	\tilde{p}_r^\mu = \frac{p_r^\mu}{1 - \alpha_K}.
\end{equation}
If the collinear set $K$ is a pair of massless particles $\set{i,j}$
the expression further simplifies to
\begin{equation}
	\alpha_{ij} = \frac{p_i\cdot p_j}
	{p_i\cdot p_j + p_i\cdot p_r + p_j\cdot p_r},
\end{equation}
which is the mapping adopted
for dipole subtraction at \gls{NLO} \cite{Catani:1996vz}.

\subsubsection{Mapping to back-to-back kinematics}
\label{ssec:catanidittmaierseymourtrocsanyi}

When the generalised rescaling mapping is applied
to exactly two momenta $p_1$ and $p_2$,
conservation laws enforce that in their centre-of-mass frame
$\vect{p}_1 + \vect{p}_2 = \vect{0}$ and
$\vect{\tilde{p}}_1 + \vect{\tilde{p}}_2 = \vect{0}$.
It is then easy to see that
\begin{align}
	\frac{E_1}{Q} = \frac{Q^2 + m_1^2 - m_2^2}{2Q^2},
	\qquad
	\frac{E_2}{Q} = \frac{Q^2 + m_2^2 - m_1^2}{2Q^2},
	\\
	\frac{\tilde{E}_1}{Q}
	= \frac{Q^2 + \tilde{m}_1^2 - \tilde{m}_2^2}{2Q^2},
	\qquad
	\frac{\tilde{E}_2}{Q}
	= \frac{Q^2 + \tilde{m}_2^2 - \tilde{m}_1^2}{2Q^2}.
\end{align}
The rescaling parameter is just the positive solution of
\begin{equation}
	\resc^2 = \frac{\lambda\brk{Q^2, m_1^2, m_2^2}}
	{\lambda\brk{Q^2, \tilde{m}_1^2, \tilde{m}_2^2}},
	\label{eq:rescpar}
\end{equation}
where $\lambda$ indicates the K\"allen function.
The full mapping reads
\begin{equation}
	\tilde{p}_1^\mu
	= \frac{1}{\resc} \brk*{p_1^\mu-\frac{E_1}{Q}Q^\mu}
	+ \frac{\tilde{E}_1}{Q}Q^\mu,
	\qquad
	\tilde{p}_2^\mu
	= \frac{1}{\resc} \brk*{p_2^\mu-\frac{E_2}{Q}Q^\mu}
	+ \frac{\tilde{E}_2}{Q}Q^\mu,
\end{equation}
and it is straightforward to see that \cref{eq:fcoll:genrescjac} reduces to
\begin{equation}
	\dd{\Phi}(Q^2; \set{m_1^2, m_2^2}) =
	\dd{\Phi}(Q^2; \set{\tilde{m}_1^2, \tilde{m}_2^2})
	\times \resc^{d-3}.
	\label{eq:fcoll:2genrescjac}
\end{equation}

In the case of one collinear cluster and one recoiler, $m_2=\tilde{m}_2$, this transformation corresponds to the momentum mapping
used to subtract quasi-collinear singularities in the dipole formalism,
when emitter and spectator are both final-state massive particles~\cite{Dittmaier:1999mb,Phaf:2001gc,Catani:2002hc}.
We have checked that \cref{eq:fcoll:2genrescjac} agrees with Refs.~\cite{Catani:1996vz} and \cite{Catani:2002hc} under this assumption.
Finally, we note that there is a unique solution for $\tilde p_1$ and $\tilde p_2$ such that they are in the $(p_1,p_2)$ plane, which here is the same as the $(p_1,Q)$ or $(p_2,Q)$ planes.
Observing that the Lorentz transformation \labelcref{eq:lorentzlambda} also ensures that the mapped momenta are in this plane,
we conclude that for one collinear cluster and one recoiler the Lorentz mapping~\cite{Nagy:2007ty}, the generalised rescaling mapping and the dipole mapping~\cite{Catani:2002hc} are all identical.

%% file: multi_collinear_map_fig.tex
\begin{figure}
\centering
\begin{tikzpicture}[baseline=0]
\begin{feynman}
	\draw (0,0) node[blob] (v) {};
	\draw[rotate= +40] (1.4,0) node {$p_1$};
	\draw[rotate= +50] (2,0) node[blob,fill=gray,scale=.4] (f) {};
	\draw[rotate= +10] (1.8,0) node {$p_2$};
	\draw[rotate= +20] (3,0) node (e) {};
	\draw[rotate= -30] (1.8,0) node {$\tilde{p}_{n-1}$};
	\draw[rotate= -20] (3,0) node (b) {};	
	\draw[rotate= -60] (1.4,0) node {$p_n$};
	\draw[rotate= -50] (2,0) node[blob,fill=gray,scale=.4] (a) {};
	\draw[rotate= -05] (2.48,0) node[dot,scale=.3] (d1) {};
	\draw[rotate= -00] (2.5,0) node[dot,scale=.3] (d2) {};
	\draw[rotate= +05] (2.48,0) node[dot,scale=.3] (d3) {};
	\draw[rotate= +55] (3.5,0) node (f1) {};
	\draw[rotate= +50] (3.5,0) node (f2) {};
	\draw[rotate= +45] (3.5,0) node (f3) {};
	\draw[rotate= -47.5] (3.5,0) node (a1) {};
	\draw[rotate= -52.5] (3.5,0) node (a2) {};
	\diagram{
		(v) -- [very thick] (f),
		(v) -- (e),
		(v) -- (b),
		(v) -- [very thick] (a),
		(a) -- (a1),
		(a) -- (a2),
		(f) -- (f1),
		(f) -- (f2),
		(f) -- (f3),
	};
\end{feynman}
\end{tikzpicture}
\hspace{3cm}
\begin{tikzpicture}[baseline=0]
\begin{feynman}
	\draw (0,0) node[blob] (v) {};
	\draw[rotate= +40] (1.8,0) node {$\tilde{p}_1$};
	\draw[rotate= +50] (3,0) node (f) {};
	\draw[rotate= +10] (1.8,0) node {$\tilde{p}_2$};
	\draw[rotate= +20] (3,0) node (e) {};
	\draw[rotate= -30] (1.8,0) node {$\tilde{p}_{n-1}$};
	\draw[rotate= -20] (3,0) node (b) {};	
	\draw[rotate= -60] (1.8,0) node {$\tilde{p}_n$};
	\draw[rotate= -50] (3,0) node (a) {};
	\draw[rotate= -05] (2.48,0) node[dot,scale=.3] (d1) {};
	\draw[rotate= -00] (2.5,0) node[dot,scale=.3] (d2) {};
	\draw[rotate= +05] (2.48,0) node[dot,scale=.3] (d3) {};
	\diagram{
		(v) -- [boson, very thick] (f),
		(v) -- (e),		
		(v) -- (b),
		(v) -- [boson, very thick] (a),
	};
\end{feynman}
\end{tikzpicture}
\caption{
Sketch of a final-collinear mapping with two simultaneous clusters of particles mapped respectively to $\tilde p_1$ and $\tilde p_n$.
We denote off-shell parent momenta with a thick line and their on-shell mapped version with a wavy line.}
\label{fig:multicol} 
\end{figure}

%% file: mapping_properties_fig.tex
\begin{figure}
\centering
\begin{subfigure}{.48\linewidth}
\centering
\scalebox{0.8}{\begin{tabular}{c c c}
\begin{tikzpicture}[baseline=0]
\begin{feynman}
	\draw (0,0) node[blob,scale=0.75] (v) {};
	\draw[rotate= +45] (1,0) node[blob,fill=gray,scale=.3] (a) {};
	\draw[rotate= +50] (2,0) node (a1) {};
	\draw[rotate= +45] (2,0) node (a2) {};
	\draw[rotate= +40] (2,0) node (a3) {};
	\draw[rotate= +40] (2.2,0) node (foo) {\( p_{123}\)};	
	\draw[rotate= 15] (1,0) node[blob,fill=gray,scale=.3] (b) {};
	\draw[rotate= +20] (2,0) node (b1) {};
	\draw[rotate= +10] (2,0) node (b2) {};
	\draw[rotate= +14] (2.2,0) node (foo) {\( p_{45}\)};		
	\draw[rotate= -15] (2,0) node (c) {};
	\draw[rotate= -45] (2,0) node (d) {};
	\diagram{
		(v) -- [very thick] (a), (a) -- (a1), (a) -- (a2), (a) -- (a3),
		(v) -- [very thick] (b), (b) -- (b1), (b) -- (b2),
		(v) -- (c),
		(v) -- (d),
	};
\end{feynman}
\end{tikzpicture}
& $\underset{p\to \tilde p}{\xrightarrow{\hspace*{2em}}}$ &
\begin{tikzpicture}[baseline=0]
\begin{feynman}
	\draw (0,0) node[blob,scale=0.75] (v) {};
	\draw[rotate= +45] (2,0) node (a) {\(\tilde p_{123}\)};
	\draw[rotate= 15] (1,0) node[blob,fill=gray,scale=.3] (b) {};
	\draw[rotate= +20] (2,0) node (b1) {};
	\draw[rotate= +10] (2,0) node (b2) {};
	\draw[rotate= +14] (2.2,0) node (foo) {\( \tilde p_{45}\)};	
	\draw[rotate= -15] (2,0) node (c) {};
	\draw[rotate= -45] (2,0) node (d) {};
	\diagram{
		(v) -- [boson, very thick] (a),
		(v) -- [very thick] (b), (b) -- (b1), (b) -- (b2),
		(v) -- (c),
		(v) -- (d),
	};
\end{feynman}
\end{tikzpicture}
\\
$\Big\downarrow {\scriptstyle p \to \hat{p}}$ & & $\Big\downarrow {\scriptstyle \tilde p \to \hat{\tilde p}}$ \\
\begin{tikzpicture}[baseline=0]
\begin{feynman}
	\draw (0,0) node[blob,scale=0.75] (v) {};
	\draw[rotate= +45] (1,0) node[blob,fill=gray,scale=.3] (a) {};
	\draw[rotate= +50] (2,0) node (a1) {};
	\draw[rotate= +45] (2,0) node (a2) {};
	\draw[rotate= +40] (2,0) node (a3) {};
	\draw[rotate= +40] (2.2,0) node (foo) {\( \hat{p}_{123}\)};
	\draw[rotate= 15] (2,0) node (b) {\(\hat{p}_{45}\)};
	\draw[rotate= -15] (2,0) node (c) {};
	\draw[rotate= -45] (2,0) node (d) {};
	\diagram{
		(v) -- [very thick] (a), (a) -- (a1), (a) -- (a2), (a) -- (a3),
		(v) -- [boson, very thick] (b),
		(v) -- (c),
		(v) -- (d),
	};
\end{feynman}
\end{tikzpicture}
& $\underset{\hat{p}\to \tilde{\hat{p}}}{\xrightarrow{\hspace*{2em}}}$ &
\begin{tikzpicture}[baseline=0]
\begin{feynman}
	\draw (0,0) node[blob,scale=0.75] (v) {};
	\draw[rotate= +45] (2,0) node (a) {\(\hat{\tilde{p}}_{123}=\tilde{\hat{p}}_{123}\)};
	\draw[rotate= 15] (2.4,0) node (b) {\(\hat{\tilde{p}}_{45}=\tilde{\hat{p}}_{45}\)};
	\draw[rotate= -15] (2,0) node (c) {};
	\draw[rotate= -45] (2,0) node (d) {};
	\diagram{
		(v) -- [boson, very thick] (a),
		(v) -- [boson, very thick] (b),
		(v) -- (c),
		(v) -- (d),
	};
\end{feynman}
\end{tikzpicture}
\end{tabular}}
\caption{Commutativity}
\label{fig:commutativity}
\end{subfigure}\vrule
\begin{subfigure}{.48\linewidth}
\centering
\scalebox{0.8}{\begin{tabular}{c c c}
\begin{tikzpicture}[baseline=0]
\begin{feynman}
	\draw (0,0) node[blob,scale=0.75] (v) {};
	\draw[rotate= +25] (0.9,0) node[blob,fill=gray,scale=.3] (a) {};
	\draw[rotate= +25] (2.4,0) node (a1) {\(p_{1}\)};
	\draw[rotate= +10] (1.5,0) node[blob,fill=gray,scale=.3] (a2) {};
	\draw[rotate= +10] (2.3,0) node (a21) {};
	\draw[rotate= +05] (2.5,0) node (fooa) {\(p_{23}\)};
	\draw[rotate= +00] (2.3,0) node (a22) {};
	\draw[rotate= -45] (2,0) node (d) {};
	\diagram{
		(v) -- [very thick] (a),
		(a) -- (a1),
		(a) -- [very thick] (a2),
		(a2) -- (a21),
		(a2) -- (a22),
		(v) -- (d),
	};
\end{feynman}
\end{tikzpicture}
&  $\underset{p\to \hat{p}}{\xrightarrow{\hspace*{2em}}}$ &
\begin{tikzpicture}[baseline=0]
\begin{feynman}
	\draw (0,0) node[blob,scale=0.75] (v) {};
	\draw[rotate= +25] (0.9,0) node[blob,fill=gray,scale=.3] (a) {};
	\draw[rotate= +25] (2.4,0) node (a1) {\(\hat{p}_{1}\)};
	\draw[rotate= +10] (2.4,0) node (a2) {\(\hat{p}_{23}\)};
	\draw[rotate= -45] (2,0) node (d) {};
	\diagram{
		(v) -- [very thick] (a),
		(a) -- (a1),
		(a) -- [boson, very thick] (a2),
		(v) -- (d),
	};
\end{feynman}
\end{tikzpicture}
\\
 & \rotatebox{-45}{$\underset{p\to \tilde{p}}{\xrightarrow{\hspace*{1.6em}}}$}  & $\Big\downarrow {\scriptstyle\hat{p}\to \bar{\hat{p}}}$ \\
& &
\begin{tikzpicture}[baseline=0]
\begin{feynman}
	\draw (0,0) node[blob,scale=0.75] (v) {};
	\draw[rotate= +25] (2.7,0) node (a) {\(\tilde{p}_{123} = \bar{\hat{p}}_{123}\)};
	\draw[rotate= -45] (2,0) node (d) {};
	\diagram{
		(v) -- [boson,very thick] (a),
		(v) -- (d),
	};
\end{feynman}
\end{tikzpicture}
\end{tabular}}
\caption{Associativity}
\label{fig:associativity}
\end{subfigure}
\caption{Schematic illustration of mapping properties. We denote massive parent momenta by a thick line splitting  and their massless mapped version by a wavy line. In the case of associativity, we show a specific iterated mapping that first groups $2$ an $3$ but any other choice yields the same final momentum if the property holds.}
\end{figure}

%% file: soft.tex
\section{Soft mappings with massive recoilers}
\label{sec:softsect}

The phase-space factorisation \labelcref{eq:PSfact_Sr},
which corresponds to the soft mapping of \cref{sec:softmappings}
does not work in the presence of massive final-state particles.
The reason is that the rescaling in \cref{eq:PS_Sr}
does not let us trade $p_i$ for $\tilde{p}_i$ in $\delta(p_i^2-m_i^2)$.
There are possible workarounds for this issue.
An example of such a fix would be to use the generalised rescaling transformation of \cref{ssec:genresc:def} to map all momenta onto the light cone, then apply the soft mapping of \cref{sec:softmappings} and finally use another generalised rescaling transformation to restore the appropriate on-shell conditions.

However, the mappings \labelcref{eq:PS_Sr,eq:PS_Smanys} also cannot be used when there is a single massive resolved particle in the final state.
Consider, for example, the real-emission corrections to inclusive Higgs production at hadron-hadron colliders via gluon fusion at \gls{NLO},
where the final state is gluon plus Higgs, $p_g + p_H$.
Then the reference vectors of the Lorentz transformation would be $K=p_H/\lambda$
and $\tilde{K}=\tilde{p}_H$.
The constraint $K^2 = \tilde{K}^2$ would imply that $\tilde{p}_H^2=p_H^2/\lambda^2$, which cannot be fulfilled by an on-shell Higgs.

A solution that lifts both issues mentioned above consists in avoiding to rescale the hard momenta in the final state, and rescale the total momentum instead.
Namely, we modify the single-soft mapping of \cref{sec:softmappings} to
$m$ partons of momenta $p_1,\ldots,p_m$ and masses $m_1,\ldots,m_m$ and one soft gluon of momentum $p_s$, 
by transforming all of the recoilers' momenta as,
\beq
\tilde{p}_i^\mu = \Lambda^\mu{}_\nu\brk[s]*{\lambda Q,Q-p_s} p_i^\nu
\quad\text{for}\quad 1\le i\le m,
\label{eq:PS_Srme}
\eeq
where $\Lambda$ is given by \cref{eq:lorentzlambda} with $\tilde{K}=\lambda_s Q$ and $K=Q-p_s$.
The constraint $K^2 = \tilde{K}^2$ is still given by \cref{eq:lambdar}, however momentum conservation becomes
\beq
	Q^\mu = p_s^\mu + \sum_{i=1}^m p_i^\mu,
	\qquad
	\lambda Q^\mu =\sum_{i=1}^m \tilde{p}_i^\mu.
\eeq
Instead of a phase space factorisation as in \cref{eq:PSfact_Sr}, we get a convolution,
\beq
\PS[m+1](\{p\}; Q=p_a+p_b)= \dd{\lambda} \PS[m](\{\tilde{p}\}_m; \lambda Q = \lambda p_a+\lambda p_b)
\: \delta\brk*{\lambda -\sqrt{1-y_{sQ}}} \frac{\dd[d]{p_s}}{(2\pi)^{d-1}}\delta_+\brk*{p_s^2}.
\label{eq:PSfact_Srme}
\eeq

Then, we extend the single-soft mapping (\ref{eq:PS_Srme}) to a multiple-soft mapping
with $m$ hard partons of momenta $p_1,\ldots,p_m$ and masses $m_1,\ldots,m_m$
and $r$ soft partons of momenta $p_{s_1},\ldots, p_{s_r}$,
\beq
\tilde{p}_i^{\mu} =
\Lambda^\mu{}_\nu\brk[s]3{\lambda_{s_1\ldots s_r} Q, Q-\sum_{j=1}^r p_{s_j}} p_i^\nu
\quad\text{for}\quad 1\le i\le m,
\label{eq:PS_Srmetrip}
\eeq
where $\Lambda$ is given in \cref{eq:lorentzlambda}, with $\ti{K}=\lambda_{s_1\ldots s_r} Q$ and $K=Q-\sum_{j=1}^r p_{s_j}$.
The constraint $K^2 = \ti{K}^2$ is given by \cref{eq:lambdamanys} which fixes $\lambda_{s_1\ldots s_r}$.
Momentum conservation becomes
\beq
Q^\mu = \sum_{j=1}^r p_{s_j}^\mu + \sum_{i=1}^m p_i^\mu, \qquad
\lambda_{s_1\ldots s_r} Q^\mu =\sum_{i=1}^m \tilde{p}_i^{\mu}.
\eeq
The phase space is given by the convolution
\beq
\PS[n](\{p\}_n;Q)= \dd{\lambda}\PS[m](\{\tilde{p}\}_m; \lambda Q)\,
\delta\brk*{\lambda-\lambda_{s_1\ldots s_r}} \brk[s]3{\prod_{j=1}^r \frac{\dd[d]{p_{s_j}}}{(2\pi)^{d-1}} \delta_+(p_{s_j}^2)}.
\label{eq:PSfact_Srtme}
\eeq

%% file: independence.tex
\section{Mapping independence of integrated counterterms}
\label{sec:independence}

Multiple momentum mappings are suitable to define counterterms that cancel the divergences of real-emission processes.
Different choice of mappings yield distinct phase-space factorisation and therefore, a priori, require redoing the integral over the unresolved degrees of freedom which yields the local cancellation of poles with virtual corrections.
The poles themselves are universal, but the finite part of the integrals depend on the choice of mapping.

In this \namecref{sec:independence}, we argue that local counterterms may be defined in such a way that the analytic form of the integrated counterterms only depends on the mapping through the integration bounds, allowing them to be used for several choices of momentum mappings.

For the sake of simplicity and concreteness, we shall discuss how this can be achieved in the case of the counterterm subtracting the divergence from a single set of $k$ final-state momenta $\left\{p\right\}_k=\left\{p_m,\dots,p_n\right\}$ becoming collinear in a $n$-particle phase space $p_1,\dots,p_n$ with $n=m+k-1$.
We refer to their parent momentum as $p_K = p_m+\dots+p_n$ and to momenta before the splitting as $\left\{p\right\}_m=\left\{p_1,\dots,p_{m-1},p_K\right\}$.
We exploit collinear factorisation of the squared amplitude to define a counterterm for this limit as follows,
\begin{align}
\abs{{\cal M}}^2\left(\set{p}_n\right) \underset{{p_m\parallel\dots \parallel p_n}}{\longrightarrow} P^{\alpha\beta}\left(\set{p}_k\right)\abs{{\cal M}}^2_{\alpha\beta}\left(\set{p}_m\right),
\end{align}
where $P^{\alpha\beta}$ is a splitting kernel, $\alpha$ and $\beta$ are spin-correlation indices, and $\abs{{\cal M}}^2_{\alpha\beta}$ is a short-hand for the spin-correlated reduced squared amplitude.
As we already discussed in section~\ref{sec:colorful_sub}, the reduced matrix element is only on shell exactly on the limit, where $p_K^2=0$, so we can define our counterterm over the full phase space as
\begin{align}
\text{CT}_{\parallel K } = P^{\alpha\beta}\left(\set{p}_k\right)\abs{{\cal M}}^2_{\alpha\beta}\left(\set{\tilde p}_m\right)\times f\left(\set{p}_m\right),
\end{align}
where $\set{\tilde p}_m$ are mapped momenta and $f$ is a function whose limit equals $1$ when $p_m\parallel\dots \parallel p_n$.

The contribution of this counterterm to the total cross section is obtained by integrating over the real-emission phase space, which we first factorise as in \cref{eq:decayfactorisation},
\begin{align}
\left\langle  \text{CT}_{\parallel K } \right\rangle &= \int \PS\left(\set{p}_n;Q\right) \text{CT}_{\parallel K }\nonumber\\
&= \int_{0}^{s_0} \frac{\dd{s_K}}{2\pi}\int \dd{\Phi_{m}}\left(\left\{ p\right\}_m;Q\right) \abs{{\cal M}}^2_{\alpha\beta}\left(\set{\tilde p}_m\right)\nonumber\\
& \times \int \dd{\Phi_{k}} \left(\set{p}_k;p_K\right)  P^{\alpha\beta}\left(\set{p}_k\right) f\left(\set{p}_n\right),
\end{align}
where $\sqrt{s_0} = Q-\sum_{i=1}^{m-1} m_i$ and the mapped momenta $\tilde p_j$ are seen as functions of the real-emission phase space.
The mapping is a change of variable $(s_K,\set{p}_m)\to (s_K,\set{\tilde p}_m) $ which transforms the mapped momenta into variables of integration and makes them independent of the variables of the splitting $s_K,p_1,\dots,p_k$,
\begin{align}
\left\langle  \text{CT}_{\parallel K } \right\rangle &= \int \dd{\tilde \Phi_m}\left(\set{\tilde p}_m;Q\right) \abs{{\cal M}}^2_{\alpha\beta}\left(\set{\tilde p}_m\right) \nonumber\\
&\times\left[\int_{0}^{\tilde s_0}  \frac{\dd{s_K}}{2\pi} J\left(s_K,\set{\tilde p}_m\right)\int \dd{\Phi_{k}} \left(\set{p}_k;p_K\right)  P^{\alpha\beta}\left(\set{p}_k\right) f\left(\set{p}_n\right)\right] \nonumber\\
&= \int \dd{\tilde\Phi_m}\left(\set{\tilde p}_m;Q\right) \abs{{\cal M}}^2_{\alpha\beta}\left(\set{\tilde p}_m\right) \left\langle P_{\alpha\beta} \right\rangle \left(\set{\tilde p}_m,\tilde s_0\right),
\label{eq:MI_CT}
\end{align}
where $\avg*{P_{\alpha\beta}}$ is the integrated kernel, $J$ is the Jacobian of the change of variable and $\tilde s_0$ is the new bound of the virtuality integral after the mapping, which can a priori be a function of the mapped phase-space variables.
This occurs, for example, in the case of the Lorentz mapping described in \cref{sec:lorentzmapping}.

As anticipated in \cref{ssec:genresc:jac}, the choice $f = J^{-1}$ appears to be particularly convenient.
In this case the condition that $f\to 1$ in the collinear limit is clearly respected, since the mapping has to become a trivial transformation in that region.
The only mapping dependence is then contained in the expression of the upper bound $\tilde s_0$ of the virtuality integral, so that computing the integral $\avg*{P_{\alpha\beta}}$ analytically as a function of $\tilde s_0$ permits the usage of the same integrated counterterm for different choices of mappings. Note that for this to be true, the integrand must be the same function of variables of $d\Phi_k$ and $d\tilde \Phi_m$ independently of the mapping, which might require adjusting the definition of splitting function variables in terms of un-mapped momenta in a mapping-dependent way.

We have verified this assertion on NLO final collinear splittings using a subtraction tool currently under development called \madnklo.
We implemented the NLO \colorful\ subtraction scheme presented in \cite{Somogyi:2009ri} within this framework, and we slightly modified it to set $f = J^{-1}$.
We validated our implementation by integrating the NLO real and virtual corrections to 3-jet production in $e^+\,e^-$ collisions and comparing to \madgraph\ \cite{Alwall:2014hca}, and we found good agreement within statistical uncertainties both globally and differentially.
We then defined a different subtraction scheme by changing the final-collinear mapping to the Lorentz mapping and setting $f$ to the appropriate inverse Jacobian factor expressed in \cref{eq:lorentzmapfactorisation}.
We kept the same functional form for the integrated counterterms as in our variation of the \colorful\ scheme and inserted the corresponding expression of $\tilde s_0$. As mentioned in the paragraph above, this implies redefining the energy fraction variable $z$ entering the splitting function such that the integrand has the same expression in terms of mapped momenta as the $z$ used in \colorful , where
\begin{align}
z_{i,ij} = \frac{p_i\cdot Q}{p_{ij}\cdot Q}\,, \qquad
\tilde{z}_{i,ij}=\frac{v(1-\alpha) \tilde{p}_{ij}\cdot Q  + \alpha Q^2/2 }{(1-\alpha) \tilde{p}_{ij}\cdot Q  + \alpha Q^2},
\end{align}
where $\alpha$ and $v$ are variables of the factorized unresolved phase space. In the \colorful\ mapping, $z_{i,ij}=\tilde{z}_{i,ij}$, but not in any other mapping. By choosing $\tilde{z}_{i,ij}$ as the energy fraction, which is a mapping dependent function of the un-mapped momenta, we ensure that the integrand of \cref{eq:MI_CT} has a mapping independent expression. 

Both the integrals over the real and virtual corrections are individually affected by this modification, but as expected their sum is not altered by the change of mapping, within sub-percent statistical uncertainties.

%% file: conclusion.tex
\section{Conclusions}
\label{sec:conclusion}

In order to obtain more and more precise theoretical predictions for the \gls{LHC} it is crucial to develop a fully-automated, efficient subtraction algorithm that can provide results at \gls{NNLO} in \gls{QCD}, and possibly beyond.
Yet ``an optimal subtraction method, able to efficiently deal with complex processes has yet to emerge''~\cite{melnikov}.
An intermediate goal would be to have a subtraction method which, up to the computation of the required two-loop amplitudes, works for every scattering process at NNLO accuracy.

While it is legitimate to push the existing subtraction methods to their maximum computational capabilities, it may be worth to dissect, analyse and compare them with the goal of eventually improving their features and components.
The work presented here was inspired by the latter point of view.

In this paper we studied momentum mappings, which are parametrisations of the phase space that factorise the variables that describe the particles becoming unresolved in some infrared or collinear limit from the variables that describe an on-shell phase space for the resolved particles.
In \cref{sec:fcoll,sec:softsect}, we have introduced new momentum mappings for final-collinear counterterms and for soft counterterms.
The new mappings work in the presence of particles of arbitrary mass and with an arbitrary number of soft particles or clusters of collinear particles, making them fit for subtraction methods at N$^k$LO accuracy, with arbitrary $k$.
In particular, the new mapping for final-collinear counterterms can also be used to show that at \gls{NLO} the mappings of Refs.~\cite{Catani:2002hc} and \cite{Nagy:2007ty} for massive particles are equivalent in the case of a single recoiler.

%% file: lorentz_transformations.tex
\section{Lorentz transformations}
\label{app:lorentztransform}

Given two Lorentz vectors $p$ and $\tilde{p}$ with $p^2 = \tilde{p}^2 \ne 0$,
a Lorentz transformation $\Lambda[\tilde{p},p]$ that maps $p$ to $\tilde{p}$
is given by \cref{eq:lorentzlambda} \cite{Catani:1996vz} which we repeat here for convenience,
\begin{equation}
\label{eq:lorentz:rotoboost}
	\Lambda^\mu{}_\nu[\tilde{p},p] =
	g^\mu{}_\nu
	- 2\frac{(p+\tilde{p})^\mu (p+\tilde{p})_\nu}{(p+\tilde{p})^2}
	+ 2\frac{\tilde{p}^\mu p_\nu}{p^2}.
\end{equation}
This expression can be obtained by writing down
the most general tensor structure that can be built out of $p$ and $\tilde{p}$,
demanding that $\Lambda[p, \tilde{p}]$ preserve the metric
and imposing $\Lambda[p, \tilde{p}]p = \tilde{p}$.
Requesting that the transformation belong
to the proper orthochronous Lorentz subgroup,
and requiring that the formula reduce to the identity for $\tilde{p} = p$
yields \cref{eq:lorentz:rotoboost} as the unique solution.
This Lorentz transformation is neither a pure boost nor a pure rotation,
but is covariant by construction.
Moreover, although by definition
$\Lambda[\tilde{p},\hat{p}] \Lambda[\hat{p}, p] p = \tilde{p}$,
in general the transformations are not associative, i.e.\ %
$\Lambda[\tilde{p},\hat{p}] \Lambda[\hat{p}, p] \ne \Lambda[\tilde{p}, p]$.
However, it can be promptly verified that the inverse operation is
$\Lambda[\tilde{p}, p]^{-1} = \Lambda[p, \tilde{p}]$.

An alternative Lorentz transformation $\Lambda[\tilde{p},p]$
that maps $p$ to $\tilde{p}$ and remains valid when $p^2 = \tilde{p}^2 = 0$ is \cite{Nagy:2007ty}
\begin{equation}
\label{eq:lorentz:boost}
	\Lambda^\mu{}_\nu[\tilde{p},p]
	= g^\mu{}_\nu
	+ \brk3{\frac{\bar{n}\cdot\tilde{p}}{\bar{n}\cdot p} - 1}
	\frac{n^\mu \bar{n}_\nu}{2}
	+ \brk3{\frac{n\cdot\tilde{p}}{n\cdot p} - 1}
	\frac{\bar{n}^\mu n_\nu}{2},
\end{equation}
$\vec{n}$ is the unit vector in the direction of $(\vec{\tilde{p}}-\vec{p})$
and the light-cone directions are defined via%
\footnote{
If $\vec{\tilde{p}} = \vec{p}$, $\vec{n}$ is ill-defined
but one may simply take $\Lambda[\tilde{p}, p]$ to be the identity.
},
\begin{equation}
	n^\mu \equiv \binom{1}{+\vec{n}},
	\qquad
	\bar{n}^\mu \equiv \binom{1}{-\vec{n}}.
\end{equation}
\Cref{eq:lorentz:boost} corresponds to a pure boost.
It can be obtained by working out the transformation laws under boosts along $\vec{n}$
of a Lorentz vector's light-cone components along the basis $\set{n, \bar{n}}$,
and imposing $\Lambda[\tilde{p}, p]p = \tilde{p}$.
We observe that when the two vectors $p$ and $\tilde{p}$ are light-like and (anti-)collinear,
a denominator in the formula will vanish.
If they are back-to-back or one of them is identically zero,
it is impossible to map $p$ into $\tilde{p}$ using a pure Lorentz boost
(although this can be achieved using general Lorentz transformations).
Otherwise it is sufficient to note that
$(n\cdot p)(\bar{n}\cdot p) = (n\cdot\tilde{p})(\bar{n}\cdot\tilde{p})$
and use the fraction that is not degenerate.
Despite its appearance, \cref{eq:lorentz:boost} is in general not covariant
due to the choice of the vector $\vec{n}$.

%% file: massive_lorentz.tex
\section{Lorentz Mapping for massive momenta}
\label{app:massive_lorentz}

In this \namecref{app:massive_lorentz}, we describe how the Lorentz mapping introduced in \cref{sec:lorentzmapping} can be extended to the case where the collinear momenta $p_{t_1},\dots, p_{t_k}$ are massive with masses $m_{t_1},\dots , m_{t_k}$ and we map $p_{t_1\dots t_k}$ to $\tilde p_{t_1\dots t_k}$ with mass $\tilde{m}_{t_1\dots t_k}$.
While the limit $p_{t_1}\parallel\dots \parallel p_{t_k}$ is not singular for nonzero masses, this generalisation can prove useful to design quasi-collinear counterterms that improve the numerical convergence in enhanced region of phase spaces.

The mapping itself has the same functional form as in the massless case, but the variables are slightly changed.
The mapped momenta are given by
\begin{align}
& \tilde p_{t_1\dots t_k}^\mu =
	\frac{1}{\lambda_{t_1\dots t_k}} \brk*{p_{t_1\dots t_k}^\mu - \frac{y_{(t_1\dots t_k)Q}}2 Q^\mu}
	+ \frac{y_{(t_1\dots t_k)Q}-\hat{y}_{t_1\dots t_k}}{2} Q^\mu,\nonumber\\
& \tilde p_i^\mu = \Lambda^\mu{}_\nu\brk[s]*{Q-\tilde p_{t_1\dots t_k}, Q-p_{t_1\dots t_k}} p_i^\nu
	\quad\text{for}\quad 1\le i \le m-1,	
\end{align}
where we have defined $\mu_i^2 \equiv m_i^2/Q^2$, $\tilde{\mu}_{t_1\dots t_k}^2 \equiv \tilde{m}_{t_1\dots t_k}^2/Q^2$,
\begin{align}
&\hat{y}_{t_1\dots t_k} = y_{t_1\dots t_k} - \tilde{\mu}_{t_1\dots t_k}^2 + \mu_{t_1}^2 + \dots + \mu_{t_k}^2,
\nonumber\\
&\lambda_{t_1\dots t_k}^2 = \frac{
	y_{(t_1\dots t_k)Q}^2 - 4\brk{\hat{y}_{t_1\dots t_k}+\tilde{\mu}_{t_1\dots t_k}^2}
}{
	\brk{y_{(t_1\dots t_k)Q}-\hat{y}_{t_1\dots t_k}}^2 - 4\tilde{\mu}_{t_1\dots t_k}^2
},
\end{align}
with $\lambda_{t_1\dots t_k}>0$.

The phase-space factorisation itself takes exactly the same functional form,
\begin{equation}
\int \dd{\Phi_{n}(\{p\}_{n};Q)} = \int \dd{\Phi_m(\{\tilde p\}_m;Q)}
\int_{s_\text{min}}^{s_0} \frac{\dd{s_{t_1\dots t_k}}}{2\pi} \lambda_{t_1\dots t_k}^{d-3}
\int \dd{\Phi_k\brk*{\set*{p_t}_k; p_{t_1\dots t_k}}},
\label{eq:lorentzmapfactorisation}
\end{equation}
where
\begin{align}
s_\text{min} &= (m_{t_1}+\dots+m_{t_k})^2 - \tilde{m}_{t_1\dots t_k}^2,\\
s_0 &= \left(1-\sqrt{1-\tilde{y}_{t_1\dots t_k}+\tilde{\mu}_{t_1\dots t_k}^2}\right)^2-\tilde{\mu}_{t_1\dots t_k}^2,\\
\tilde{y}_{t_1\dots t_k}&= \frac{2 p_{t_1\dots t_k}\cdot Q}{Q^2}
\end{align}

%% file: main.bbl
\providecommand{\href}[2]{#2}\begingroup\raggedright\begin{thebibliography}{10}

\bibitem{Aad:2012tfa}
{\scshape ATLAS} collaboration, \emph{{Observation of a new particle in the
  search for the Standard Model Higgs boson with the ATLAS detector at the
  LHC}}, \href{https://doi.org/10.1016/j.physletb.2012.08.020}{\emph{Phys.
  Lett.} {\bfseries B716} (2012) 1}
  [\href{https://arxiv.org/abs/1207.7214}{{\ttfamily 1207.7214}}].

\bibitem{Chatrchyan:2012xdj}
{\scshape CMS} collaboration, \emph{{Observation of a new boson at a mass of
  125 GeV with the CMS experiment at the LHC}},
  \href{https://doi.org/10.1016/j.physletb.2012.08.021}{\emph{Phys. Lett.}
  {\bfseries B716} (2012) 30}
  [\href{https://arxiv.org/abs/1207.7235}{{\ttfamily 1207.7235}}].

\bibitem{Kinoshita:1962ur}
T.~Kinoshita, \emph{{Mass singularities of Feynman amplitudes}},
  \href{https://doi.org/10.1063/1.1724268}{\emph{J. Math. Phys.} {\bfseries 3}
  (1962) 650}.

\bibitem{Lee:1964is}
T.~D. Lee and M.~Nauenberg, \emph{{Degenerate Systems and Mass Singularities}},
  \href{https://doi.org/10.1103/PhysRev.133.B1549}{\emph{Phys. Rev.} {\bfseries
  133} (1964) B1549}.

\bibitem{Catani:1998bh}
S.~Catani, \emph{{The Singular behavior of QCD amplitudes at two loop order}},
  \href{https://doi.org/10.1016/S0370-2693(98)00332-3}{\emph{Phys. Lett.}
  {\bfseries B427} (1998) 161}
  [\href{https://arxiv.org/abs/hep-ph/9802439}{{\ttfamily hep-ph/9802439}}].

\bibitem{Sterman:2002qn}
G.~F. Sterman and M.~E. Tejeda-Yeomans, \emph{{Multiloop amplitudes and
  resummation}},
  \href{https://doi.org/10.1016/S0370-2693(02)03100-3}{\emph{Phys. Lett.}
  {\bfseries B552} (2003) 48}
  [\href{https://arxiv.org/abs/hep-ph/0210130}{{\ttfamily hep-ph/0210130}}].

\bibitem{Aybat:2006wq}
S.~M. Aybat, L.~J. Dixon and G.~F. Sterman, \emph{{The Two-loop anomalous
  dimension matrix for soft gluon exchange}},
  \href{https://doi.org/10.1103/PhysRevLett.97.072001}{\emph{Phys. Rev. Lett.}
  {\bfseries 97} (2006) 072001}
  [\href{https://arxiv.org/abs/hep-ph/0606254}{{\ttfamily hep-ph/0606254}}].

\bibitem{Aybat:2006mz}
S.~M. Aybat, L.~J. Dixon and G.~F. Sterman, \emph{{The Two-loop soft anomalous
  dimension matrix and resummation at next-to-next-to leading pole}},
  \href{https://doi.org/10.1103/PhysRevD.74.074004}{\emph{Phys. Rev.}
  {\bfseries D74} (2006) 074004}
  [\href{https://arxiv.org/abs/hep-ph/0607309}{{\ttfamily hep-ph/0607309}}].

\bibitem{Bern:1994zx}
Z.~Bern, L.~J. Dixon, D.~C. Dunbar and D.~A. Kosower, \emph{{One loop n point
  gauge theory amplitudes, unitarity and collinear limits}},
  \href{https://doi.org/10.1016/0550-3213(94)90179-1}{\emph{Nucl. Phys.}
  {\bfseries B425} (1994) 217}
  [\href{https://arxiv.org/abs/hep-ph/9403226}{{\ttfamily hep-ph/9403226}}].

\bibitem{Bern:1998sc}
Z.~Bern, V.~Del~Duca and C.~R. Schmidt, \emph{{The Infrared behavior of one
  loop gluon amplitudes at next-to-next-to-leading order}},
  \href{https://doi.org/10.1016/S0370-2693(98)01495-6}{\emph{Phys. Lett.}
  {\bfseries B445} (1998) 168}
  [\href{https://arxiv.org/abs/hep-ph/9810409}{{\ttfamily hep-ph/9810409}}].

\bibitem{Kosower:1999rx}
D.~A. Kosower and P.~Uwer, \emph{{One loop splitting amplitudes in gauge
  theory}}, \href{https://doi.org/10.1016/S0550-3213(99)00583-0}{\emph{Nucl.
  Phys.} {\bfseries B563} (1999) 477}
  [\href{https://arxiv.org/abs/hep-ph/9903515}{{\ttfamily hep-ph/9903515}}].

\bibitem{Bern:1999ry}
Z.~Bern, V.~Del~Duca, W.~B. Kilgore and C.~R. Schmidt, \emph{{The infrared
  behavior of one loop QCD amplitudes at next-to-next-to leading order}},
  \href{https://doi.org/10.1103/PhysRevD.60.116001}{\emph{Phys. Rev.}
  {\bfseries D60} (1999) 116001}
  [\href{https://arxiv.org/abs/hep-ph/9903516}{{\ttfamily hep-ph/9903516}}].

\bibitem{Catani:2000pi}
S.~Catani and M.~Grazzini, \emph{{The soft gluon current at one loop order}},
  \href{https://doi.org/10.1016/S0550-3213(00)00572-1}{\emph{Nucl. Phys.}
  {\bfseries B591} (2000) 435}
  [\href{https://arxiv.org/abs/hep-ph/0007142}{{\ttfamily hep-ph/0007142}}].

\bibitem{Campbell:1997hg}
J.~M. Campbell and E.~W.~N. Glover, \emph{{Double unresolved approximations to
  multiparton scattering amplitudes}},
  \href{https://doi.org/10.1016/S0550-3213(98)00295-8}{\emph{Nucl. Phys.}
  {\bfseries B527} (1998) 264}
  [\href{https://arxiv.org/abs/hep-ph/9710255}{{\ttfamily hep-ph/9710255}}].

\bibitem{Catani:1999ss}
S.~Catani and M.~Grazzini, \emph{{Infrared factorization of tree level QCD
  amplitudes at the next-to-next-to-leading order and beyond}},
  \href{https://doi.org/10.1016/S0550-3213(99)00778-6}{\emph{Nucl. Phys.}
  {\bfseries B570} (2000) 287}
  [\href{https://arxiv.org/abs/hep-ph/9908523}{{\ttfamily hep-ph/9908523}}].

\bibitem{DelDuca:1999iql}
V.~Del~Duca, A.~Frizzo and F.~Maltoni, \emph{{Factorization of tree QCD
  amplitudes in the high-energy limit and in the collinear limit}},
  \href{https://doi.org/10.1016/S0550-3213(99)00657-4}{\emph{Nucl. Phys.}
  {\bfseries B568} (2000) 211}
  [\href{https://arxiv.org/abs/hep-ph/9909464}{{\ttfamily hep-ph/9909464}}].

\bibitem{Kosower:2002su}
D.~A. Kosower, \emph{{Multiple singular emission in gauge theories}},
  \href{https://doi.org/10.1103/PhysRevD.67.116003}{\emph{Phys. Rev.}
  {\bfseries D67} (2003) 116003}
  [\href{https://arxiv.org/abs/hep-ph/0212097}{{\ttfamily hep-ph/0212097}}].

\bibitem{Almelid:2015jia}
O.~Almelid, C.~Duhr and E.~Gardi, \emph{{Three-loop corrections to the soft
  anomalous dimension in multileg scattering}},
  \href{https://doi.org/10.1103/PhysRevLett.117.172002}{\emph{Phys. Rev. Lett.}
  {\bfseries 117} (2016) 172002}
  [\href{https://arxiv.org/abs/1507.00047}{{\ttfamily 1507.00047}}].

\bibitem{Bern:2004cz}
Z.~Bern, L.~J. Dixon and D.~A. Kosower, \emph{{Two-loop g ---> gg splitting
  amplitudes in QCD}},
  \href{https://doi.org/10.1088/1126-6708/2004/08/012}{\emph{JHEP} {\bfseries
  08} (2004) 012} [\href{https://arxiv.org/abs/hep-ph/0404293}{{\ttfamily
  hep-ph/0404293}}].

\bibitem{Badger:2004uk}
S.~D. Badger and E.~W.~N. Glover, \emph{{Two loop splitting functions in QCD}},
  \href{https://doi.org/10.1088/1126-6708/2004/07/040}{\emph{JHEP} {\bfseries
  07} (2004) 040} [\href{https://arxiv.org/abs/hep-ph/0405236}{{\ttfamily
  hep-ph/0405236}}].

\bibitem{Duhr:2013msa}
C.~Duhr and T.~Gehrmann, \emph{{The two-loop soft current in dimensional
  regularization}},
  \href{https://doi.org/10.1016/j.physletb.2013.10.063}{\emph{Phys. Lett.}
  {\bfseries B727} (2013) 452}
  [\href{https://arxiv.org/abs/1309.4393}{{\ttfamily 1309.4393}}].

\bibitem{Li:2013lsa}
Y.~Li and H.~X. Zhu, \emph{{Single soft gluon emission at two loops}},
  \href{https://doi.org/10.1007/JHEP11(2013)080}{\emph{JHEP} {\bfseries 11}
  (2013) 080} [\href{https://arxiv.org/abs/1309.4391}{{\ttfamily 1309.4391}}].

\bibitem{Duhr:2014nda}
C.~Duhr, T.~Gehrmann and M.~Jaquier, \emph{{Two-loop splitting amplitudes and
  the single-real contribution to inclusive Higgs production at N$^3$LO}},
  \href{https://doi.org/10.1007/JHEP02(2015)077}{\emph{JHEP} {\bfseries 02}
  (2015) 077} [\href{https://arxiv.org/abs/1411.3587}{{\ttfamily 1411.3587}}].

\bibitem{Catani:2003vu}
S.~Catani, D.~de~Florian and G.~Rodrigo, \emph{{The Triple collinear limit of
  one loop QCD amplitudes}},
  \href{https://doi.org/10.1016/j.physletb.2004.02.039}{\emph{Phys. Lett.}
  {\bfseries B586} (2004) 323}
  [\href{https://arxiv.org/abs/hep-ph/0312067}{{\ttfamily hep-ph/0312067}}].

\bibitem{Badger:2015cxa}
S.~Badger, F.~Buciuni and T.~Peraro, \emph{{One-loop triple collinear splitting
  amplitudes in QCD}},
  \href{https://doi.org/10.1007/JHEP09(2015)188}{\emph{JHEP} {\bfseries 09}
  (2015) 188} [\href{https://arxiv.org/abs/1507.05070}{{\ttfamily
  1507.05070}}].

\bibitem{Birthwright:2005ak}
T.~G. Birthwright, E.~W.~N. Glover, V.~V. Khoze and P.~Marquard,
  \emph{{Multi-gluon collinear limits from MHV diagrams}},
  \href{https://doi.org/10.1088/1126-6708/2005/05/013}{\emph{JHEP} {\bfseries
  05} (2005) 013} [\href{https://arxiv.org/abs/hep-ph/0503063}{{\ttfamily
  hep-ph/0503063}}].

\bibitem{Duhr:2006}
C.~Duhr, \emph{{Applications of twistor methods in QCD}},  Master's thesis,
  Universit\'e catholique de Louvain, 2006.

\bibitem{Catani:2019nqv}
S.~Catani, D.~Colferai and A.~Torrini, \emph{{Triple (and quadruple) soft-gluon
  radiation in QCD hard scattering}},
  \href{https://arxiv.org/abs/1908.01616}{{\ttfamily 1908.01616}}.

\bibitem{Frixione:1995ms}
S.~Frixione, Z.~Kunszt and A.~Signer, \emph{{Three jet cross-sections to
  next-to-leading order}},
  \href{https://doi.org/10.1016/0550-3213(96)00110-1}{\emph{Nucl. Phys.}
  {\bfseries B467} (1996) 399}
  [\href{https://arxiv.org/abs/hep-ph/9512328}{{\ttfamily hep-ph/9512328}}].

\bibitem{Catani:1996vz}
S.~Catani and M.~H. Seymour, \emph{{A General algorithm for calculating jet
  cross-sections in NLO QCD}},
  \href{https://doi.org/10.1016/S0550-3213(96)00589-5,
  10.1016/S0550-3213(98)81022-5}{\emph{Nucl. Phys.} {\bfseries B485} (1997)
  291} [\href{https://arxiv.org/abs/hep-ph/9605323}{{\ttfamily
  hep-ph/9605323}}].

\bibitem{Catani:2007vq}
S.~Catani and M.~Grazzini, \emph{{An NNLO subtraction formalism in hadron
  collisions and its application to Higgs boson production at the LHC}},
  \href{https://doi.org/10.1103/PhysRevLett.98.222002}{\emph{Phys. Rev. Lett.}
  {\bfseries 98} (2007) 222002}
  [\href{https://arxiv.org/abs/hep-ph/0703012}{{\ttfamily hep-ph/0703012}}].

\bibitem{Boughezal:2015dva}
R.~Boughezal, C.~Focke, X.~Liu and F.~Petriello, \emph{{$W$-boson production in
  association with a jet at next-to-next-to-leading order in perturbative
  QCD}}, \href{https://doi.org/10.1103/PhysRevLett.115.062002}{\emph{Phys. Rev.
  Lett.} {\bfseries 115} (2015) 062002}
  [\href{https://arxiv.org/abs/1504.02131}{{\ttfamily 1504.02131}}].

\bibitem{Gaunt:2015pea}
J.~Gaunt, M.~Stahlhofen, F.~J. Tackmann and J.~R. Walsh, \emph{{N-jettiness
  Subtractions for NNLO QCD Calculations}},
  \href{https://doi.org/10.1007/JHEP09(2015)058}{\emph{JHEP} {\bfseries 09}
  (2015) 058} [\href{https://arxiv.org/abs/1505.04794}{{\ttfamily
  1505.04794}}].

\bibitem{GehrmannDeRidder:2005cm}
A.~Gehrmann-De~Ridder, T.~Gehrmann and E.~W.~N. Glover, \emph{{Antenna
  subtraction at NNLO}},
  \href{https://doi.org/10.1088/1126-6708/2005/09/056}{\emph{JHEP} {\bfseries
  09} (2005) 056} [\href{https://arxiv.org/abs/hep-ph/0505111}{{\ttfamily
  hep-ph/0505111}}].

\bibitem{Daleo:2006xa}
A.~Daleo, T.~Gehrmann and D.~Maitre, \emph{{Antenna subtraction with hadronic
  initial states}},
  \href{https://doi.org/10.1088/1126-6708/2007/04/016}{\emph{JHEP} {\bfseries
  04} (2007) 016} [\href{https://arxiv.org/abs/hep-ph/0612257}{{\ttfamily
  hep-ph/0612257}}].

\bibitem{GehrmannDeRidder:2005aw}
A.~Gehrmann-De~Ridder, T.~Gehrmann and E.~W.~N. Glover, \emph{{Gluon-gluon
  antenna functions from Higgs boson decay}},
  \href{https://doi.org/10.1016/j.physletb.2005.03.003}{\emph{Phys. Lett.}
  {\bfseries B612} (2005) 49}
  [\href{https://arxiv.org/abs/hep-ph/0502110}{{\ttfamily hep-ph/0502110}}].

\bibitem{GehrmannDeRidder:2005hi}
A.~Gehrmann-De~Ridder, T.~Gehrmann and E.~W.~N. Glover, \emph{{Quark-gluon
  antenna functions from neutralino decay}},
  \href{https://doi.org/10.1016/j.physletb.2005.02.039}{\emph{Phys. Lett.}
  {\bfseries B612} (2005) 36}
  [\href{https://arxiv.org/abs/hep-ph/0501291}{{\ttfamily hep-ph/0501291}}].

\bibitem{Daleo:2009yj}
A.~Daleo, A.~Gehrmann-De~Ridder, T.~Gehrmann and G.~Luisoni, \emph{{Antenna
  subtraction at NNLO with hadronic initial states: initial-final
  configurations}}, \href{https://doi.org/10.1007/JHEP01(2010)118}{\emph{JHEP}
  {\bfseries 01} (2010) 118} [\href{https://arxiv.org/abs/0912.0374}{{\ttfamily
  0912.0374}}].

\bibitem{Gehrmann:2011wi}
T.~Gehrmann and P.~F. Monni, \emph{{Antenna subtraction at NNLO with hadronic
  initial states: real-virtual initial-initial configurations}},
  \href{https://doi.org/10.1007/JHEP12(2011)049}{\emph{JHEP} {\bfseries 12}
  (2011) 049} [\href{https://arxiv.org/abs/1107.4037}{{\ttfamily 1107.4037}}].

\bibitem{Boughezal:2010mc}
R.~Boughezal, A.~Gehrmann-De~Ridder and M.~Ritzmann, \emph{{Antenna subtraction
  at NNLO with hadronic initial states: double real radiation for
  initial-initial configurations with two quark flavours}},
  \href{https://doi.org/10.1007/JHEP02(2011)098}{\emph{JHEP} {\bfseries 02}
  (2011) 098} [\href{https://arxiv.org/abs/1011.6631}{{\ttfamily 1011.6631}}].

\bibitem{GehrmannDeRidder:2012ja}
A.~Gehrmann-De~Ridder, T.~Gehrmann and M.~Ritzmann, \emph{{Antenna subtraction
  at NNLO with hadronic initial states: double real initial-initial
  configurations}}, \href{https://doi.org/10.1007/JHEP10(2012)047}{\emph{JHEP}
  {\bfseries 10} (2012) 047} [\href{https://arxiv.org/abs/1207.5779}{{\ttfamily
  1207.5779}}].

\bibitem{Currie:2013vh}
J.~Currie, E.~W.~N. Glover and S.~Wells, \emph{{Infrared Structure at NNLO
  Using Antenna Subtraction}},
  \href{https://doi.org/10.1007/JHEP04(2013)066}{\emph{JHEP} {\bfseries 04}
  (2013) 066} [\href{https://arxiv.org/abs/1301.4693}{{\ttfamily 1301.4693}}].

\bibitem{Somogyi:2005xz}
G.~Somogyi, Z.~Trocsanyi and V.~Del~Duca, \emph{{Matching of singly- and
  doubly-unresolved limits of tree-level QCD squared matrix elements}},
  \href{https://doi.org/10.1088/1126-6708/2005/06/024}{\emph{JHEP} {\bfseries
  06} (2005) 024} [\href{https://arxiv.org/abs/hep-ph/0502226}{{\ttfamily
  hep-ph/0502226}}].

\bibitem{Somogyi:2006da}
G.~Somogyi, Z.~Trocsanyi and V.~Del~Duca, \emph{{A Subtraction scheme for
  computing QCD jet cross sections at NNLO: Regularization of doubly-real
  emissions}}, \href{https://doi.org/10.1088/1126-6708/2007/01/070}{\emph{JHEP}
  {\bfseries 01} (2007) 070}
  [\href{https://arxiv.org/abs/hep-ph/0609042}{{\ttfamily hep-ph/0609042}}].

\bibitem{Somogyi:2006db}
G.~Somogyi and Z.~Trocsanyi, \emph{{A Subtraction scheme for computing QCD jet
  cross sections at NNLO: Regularization of real-virtual emission}},
  \href{https://doi.org/10.1088/1126-6708/2007/01/052}{\emph{JHEP} {\bfseries
  01} (2007) 052} [\href{https://arxiv.org/abs/hep-ph/0609043}{{\ttfamily
  hep-ph/0609043}}].

\bibitem{Somogyi:2008fc}
G.~Somogyi and Z.~Trocsanyi, \emph{{A Subtraction scheme for computing QCD jet
  cross sections at NNLO: Integrating the subtraction terms. I.}},
  \href{https://doi.org/10.1088/1126-6708/2008/08/042}{\emph{JHEP} {\bfseries
  08} (2008) 042} [\href{https://arxiv.org/abs/0807.0509}{{\ttfamily
  0807.0509}}].

\bibitem{Aglietti:2008fe}
U.~Aglietti, V.~Del~Duca, C.~Duhr, G.~Somogyi and Z.~Trocsanyi, \emph{{Analytic
  integration of real-virtual counterterms in NNLO jet cross sections. I.}},
  \href{https://doi.org/10.1088/1126-6708/2008/09/107}{\emph{JHEP} {\bfseries
  09} (2008) 107} [\href{https://arxiv.org/abs/0807.0514}{{\ttfamily
  0807.0514}}].

\bibitem{Somogyi:2009ri}
G.~Somogyi, \emph{{Subtraction with hadronic initial states at NLO: An
  NNLO-compatible scheme}},
  \href{https://doi.org/10.1088/1126-6708/2009/05/016}{\emph{JHEP} {\bfseries
  05} (2009) 016} [\href{https://arxiv.org/abs/0903.1218}{{\ttfamily
  0903.1218}}].

\bibitem{Bolzoni:2009ye}
P.~Bolzoni, S.-O. Moch, G.~Somogyi and Z.~Trocsanyi, \emph{{Analytic
  integration of real-virtual counterterms in NNLO jet cross sections. II.}},
  \href{https://doi.org/10.1088/1126-6708/2009/08/079}{\emph{JHEP} {\bfseries
  08} (2009) 079} [\href{https://arxiv.org/abs/0905.4390}{{\ttfamily
  0905.4390}}].

\bibitem{Bolzoni:2010bt}
P.~Bolzoni, G.~Somogyi and Z.~Trocsanyi, \emph{{A subtraction scheme for
  computing QCD jet cross sections at NNLO: integrating the iterated
  singly-unresolved subtraction terms}},
  \href{https://doi.org/10.1007/JHEP01(2011)059}{\emph{JHEP} {\bfseries 01}
  (2011) 059} [\href{https://arxiv.org/abs/1011.1909}{{\ttfamily 1011.1909}}].

\bibitem{DelDuca:2013kw}
V.~Del~Duca, G.~Somogyi and Z.~Trocsanyi, \emph{{Integration of collinear-type
  doubly unresolved counterterms in NNLO jet cross sections}},
  \href{https://doi.org/10.1007/JHEP06(2013)079}{\emph{JHEP} {\bfseries 06}
  (2013) 079} [\href{https://arxiv.org/abs/1301.3504}{{\ttfamily 1301.3504}}].

\bibitem{Somogyi:2013yk}
G.~Somogyi, \emph{{A subtraction scheme for computing QCD jet cross sections at
  NNLO: integrating the doubly unresolved subtraction terms}},
  \href{https://doi.org/10.1007/JHEP04(2013)010}{\emph{JHEP} {\bfseries 04}
  (2013) 010} [\href{https://arxiv.org/abs/1301.3919}{{\ttfamily 1301.3919}}].

\bibitem{Czakon:2010td}
M.~Czakon, \emph{{A novel subtraction scheme for double-real radiation at
  NNLO}}, \href{https://doi.org/10.1016/j.physletb.2010.08.036}{\emph{Phys.
  Lett.} {\bfseries B693} (2010) 259}
  [\href{https://arxiv.org/abs/1005.0274}{{\ttfamily 1005.0274}}].

\bibitem{Czakon:2011ve}
M.~Czakon, \emph{{Double-real radiation in hadronic top quark pair production
  as a proof of a certain concept}},
  \href{https://doi.org/10.1016/j.nuclphysb.2011.03.020}{\emph{Nucl. Phys.}
  {\bfseries B849} (2011) 250}
  [\href{https://arxiv.org/abs/1101.0642}{{\ttfamily 1101.0642}}].

\bibitem{Czakon:2014oma}
M.~Czakon and D.~Heymes, \emph{{Four-dimensional formulation of the
  sector-improved residue subtraction scheme}},
  \href{https://doi.org/10.1016/j.nuclphysb.2014.11.006}{\emph{Nucl. Phys.}
  {\bfseries B890} (2014) 152}
  [\href{https://arxiv.org/abs/1408.2500}{{\ttfamily 1408.2500}}].

\bibitem{Czakon:2019tmo}
M.~Czakon, A.~van Hameren, A.~Mitov and R.~Poncelet, \emph{{Single-jet
  inclusive rates with exact color at $\mathcal{O}(\alpha_s^4)$}},
  \href{https://arxiv.org/abs/1907.12911}{{\ttfamily 1907.12911}}.

\bibitem{Caola:2017dug}
F.~Caola, K.~Melnikov and R.~Röntsch, \emph{{Nested soft-collinear
  subtractions in NNLO QCD computations}},
  \href{https://doi.org/10.1140/epjc/s10052-017-4774-0}{\emph{Eur. Phys. J.}
  {\bfseries C77} (2017) 248}
  [\href{https://arxiv.org/abs/1702.01352}{{\ttfamily 1702.01352}}].

\bibitem{Caola:2018pxp}
F.~Caola, M.~Delto, H.~Frellesvig and K.~Melnikov, \emph{{The double-soft
  integral for an arbitrary angle between hard radiators}},
  \href{https://doi.org/10.1140/epjc/s10052-018-6180-7}{\emph{Eur. Phys. J.}
  {\bfseries C78} (2018) 687}
  [\href{https://arxiv.org/abs/1807.05835}{{\ttfamily 1807.05835}}].

\bibitem{Delto:2019asp}
M.~Delto and K.~Melnikov, \emph{{Integrated triple-collinear counter-terms for
  the nested soft-collinear subtraction scheme}},
  \href{https://doi.org/10.1007/JHEP05(2019)148}{\emph{JHEP} {\bfseries 05}
  (2019) 148} [\href{https://arxiv.org/abs/1901.05213}{{\ttfamily
  1901.05213}}].

\bibitem{Caola:2019nzf}
F.~Caola, K.~Melnikov and R.~Röntsch, \emph{{Analytic results for
  color-singlet production at NNLO QCD with the nested soft-collinear
  subtraction scheme}},  \href{https://arxiv.org/abs/1902.02081}{{\ttfamily
  1902.02081}}.

\bibitem{Caola:2019pfz}
F.~Caola, K.~Melnikov and R.~Röntsch, \emph{{Analytic results for decays of
  color singlets to $gg$ and $q \bar q$ final states at NNLO QCD with the
  nested soft-collinear subtraction scheme}},
  \href{https://arxiv.org/abs/1907.05398}{{\ttfamily 1907.05398}}.

\bibitem{Cacciari:2015jma}
M.~Cacciari, F.~A. Dreyer, A.~Karlberg, G.~P. Salam and G.~Zanderighi,
  \emph{{Fully Differential Vector-Boson-Fusion Higgs Production at
  Next-to-Next-to-Leading Order}},
  \href{https://doi.org/10.1103/PhysRevLett.115.082002,
  10.1103/PhysRevLett.120.139901}{\emph{Phys. Rev. Lett.} {\bfseries 115}
  (2015) 082002} [\href{https://arxiv.org/abs/1506.02660}{{\ttfamily
  1506.02660}}].

\bibitem{Magnea:2018hab}
L.~Magnea, E.~Maina, G.~Pelliccioli, C.~Signorile-Signorile, P.~Torrielli and
  S.~Uccirati, \emph{{Local Analytic Sector Subtraction at NNLO}},
  \href{https://doi.org/10.1007/JHEP12(2018)107}{\emph{JHEP} {\bfseries 12}
  (2018) 107} [\href{https://arxiv.org/abs/1806.09570}{{\ttfamily
  1806.09570}}].

\bibitem{Magnea:2018ebr}
L.~Magnea, E.~Maina, G.~Pelliccioli, C.~Signorile-Signorile, P.~Torrielli and
  S.~Uccirati, \emph{{Factorisation and Subtraction beyond NLO}},
  \href{https://doi.org/10.1007/JHEP12(2018)062}{\emph{JHEP} {\bfseries 12}
  (2018) 062} [\href{https://arxiv.org/abs/1809.05444}{{\ttfamily
  1809.05444}}].

\bibitem{Herzog:2018ily}
F.~Herzog, \emph{{Geometric IR subtraction for final state real radiation}},
  \href{https://doi.org/10.1007/JHEP08(2018)006}{\emph{JHEP} {\bfseries 08}
  (2018) 006} [\href{https://arxiv.org/abs/1804.07949}{{\ttfamily
  1804.07949}}].

\bibitem{Somogyi:2006cz}
G.~Somogyi and Z.~Trocsanyi, \emph{{A New subtraction scheme for computing QCD
  jet cross sections at next-to-leading order accuracy}},
  \href{https://arxiv.org/abs/hep-ph/0609041}{{\ttfamily hep-ph/0609041}}.

\bibitem{Catani:2002hc}
S.~Catani, S.~Dittmaier, M.~H. Seymour and Z.~Trocsanyi, \emph{{The Dipole
  formalism for next-to-leading order QCD calculations with massive partons}},
  \href{https://doi.org/10.1016/S0550-3213(02)00098-6}{\emph{Nucl. Phys.}
  {\bfseries B627} (2002) 189}
  [\href{https://arxiv.org/abs/hep-ph/0201036}{{\ttfamily hep-ph/0201036}}].

\bibitem{Nagy:2007ty}
Z.~Nagy and D.~E. Soper, \emph{{Parton showers with quantum interference}},
  \href{https://doi.org/10.1088/1126-6708/2007/09/114}{\emph{JHEP} {\bfseries
  09} (2007) 114} [\href{https://arxiv.org/abs/0706.0017}{{\ttfamily
  0706.0017}}].

\bibitem{Gehrmann-DeRidder:2003pne}
A.~Gehrmann-De~Ridder, T.~Gehrmann and G.~Heinrich, \emph{{Four particle phase
  space integrals in massless QCD}},
  \href{https://doi.org/10.1016/j.nuclphysb.2004.01.023}{\emph{Nucl. Phys.}
  {\bfseries B682} (2004) 265}
  [\href{https://arxiv.org/abs/hep-ph/0311276}{{\ttfamily hep-ph/0311276}}].

\bibitem{Dittmaier:1999mb}
S.~Dittmaier, \emph{{A General approach to photon radiation off fermions}},
  \href{https://doi.org/10.1016/S0550-3213(99)00563-5}{\emph{Nucl. Phys.}
  {\bfseries B565} (2000) 69}
  [\href{https://arxiv.org/abs/hep-ph/9904440}{{\ttfamily hep-ph/9904440}}].

\bibitem{Phaf:2001gc}
L.~Phaf and S.~Weinzierl, \emph{{Dipole formalism with heavy fermions}},
  \href{https://doi.org/10.1088/1126-6708/2001/04/006}{\emph{JHEP} {\bfseries
  04} (2001) 006} [\href{https://arxiv.org/abs/hep-ph/0102207}{{\ttfamily
  hep-ph/0102207}}].

\bibitem{Alwall:2014hca}
J.~Alwall, R.~Frederix, S.~Frixione, V.~Hirschi, F.~Maltoni, O.~Mattelaer
  et~al., \emph{{The automated computation of tree-level and next-to-leading
  order differential cross sections, and their matching to parton shower
  simulations}}, \href{https://doi.org/10.1007/JHEP07(2014)079}{\emph{JHEP}
  {\bfseries 07} (2014) 079} [\href{https://arxiv.org/abs/1405.0301}{{\ttfamily
  1405.0301}}].

\bibitem{melnikov}
K.~Melnikov, \emph{{Precision Physics at the LHC: what and how}},  in
  \emph{Amplitudes}, Dublin 2019.

\end{thebibliography}\endgroup
